\documentclass[twocolumn]{aastex63}

\usepackage{lmodern}
\usepackage{latexsym}
\usepackage{amsmath}
\usepackage{amssymb}
\usepackage{amsbsy}
\usepackage{amsthm}
\usepackage{amsfonts}
\usepackage{mathrsfs}
\usepackage{bm}
\usepackage{sansmath}
\usepackage{relsize}
\usepackage{caption2}
\usepackage{graphicx}

\submitjournal{ApJ}

\shorttitle{outflow from supercritical accretion flow}
\shortauthors{Zeraatgari et al.}

\begin{document}

\title{Two-dimensional Inflow-outflow Solution of Supercritical Accretion Flow}

\correspondingauthor{Amin Mosallanezhad}
\email{mosallanezhad@ustc.edu.cn}

\author[0000-0003-3345-727X]{Fatemeh Zahra Zeraatgari}
\affiliation{School of Mathematics and Statistics, Xi'an Jiaotong University, Xi'an, Shaanxi 710049, PR China}

\author[0000-0002-4601-7073]{Amin Mosallanezhad}
\affiliation{Key Laboratory for Research in Galaxies and Cosmology, Department of Astronomy,\\ 
University of Science and Technology of China, Hefei, Anhui 230026, PR China}

\author[0000-0002-7330-4756]{Ye-Fei Yuan}
\affiliation{Key Laboratory for Research in Galaxies and Cosmology, Department of Astronomy,\\ 
University of Science and Technology of China, Hefei, Anhui 230026, PR China}

\author[0000-0002-0427-520X]{De-Fu Bu}
\affiliation{Key Laboratory for Research in Galaxies and Cosmology, Shanghai Astronomical Observatory,\\
Chinese Academy of Sciences, 80 Nandan Road, Shanghai 200030, PR China}

\author[0000-0003-3468-8803]{Liquan Mei}
\affiliation{School of Mathematics and Statistics, Xi'an Jiaotong University, Xi'an, Shaanxi 710049, PR China}



\begin{abstract}

We present the two-dimensional inflow-outflow solutions of radiation hydrodynamic (RHD) 
equations of supercritical accretion flows. Compared with prior studies, we include all 
components of the viscous stress tensor. We assume steady state flow and use self-similar 
solutions in the radial direction to solve the equations in $ r-\theta $ domain of the spherical 
coordinates. The set of differential equations have been integrated from the rotation axis to 
the equatorial plane. We find that the self-similarity assumption requires that the 
radial profile of density is described by $ \rho(r) \propto r^{-0.5} $. Correspondingly, the radial profile of the mass inflow 
rate decreases with decreasing radii as $ \dot{M}_\mathrm{in} \propto r $. Inflow-outflow structure
has been found in our solution. In the region $ \theta > 65^{\circ} $ there exist inflow while above 
that flow moves outward and outflow could launch. The driving forces of the outflow are 
analyzed and found that the radiation force is dominant and push the gas particles outwards with 
poloidal velocity $ \sim 0.25 c $. The properties of outflow are also studied. The results 
show that the mass flux weighted angular momentum of the inflow is lower than that of 
outflow, thus the angular momentum of the flow can be transported by the outflow.
We also analyze the convective stability of the supercritical disk and find that in the absence of 
the magnetic field, the flow is convectively unstable. Our analytical results are fully consistent with 
the previous numerical simulations of the supercritical accretion flow.

\end{abstract}



\section{Introduction} \label{sec:intro}

Accretion of gas through a disc onto a black hole is associated with 
many active phenomena in our universe such as active galactic nuclei
(AGNs), X-ray binaries (XRBs), and extra galactic jets. 
Based on temperature, black hole accretion discs can be divided into two
distinct classes: hot and cold (see \citealt{Yuan and Narayan 2014} for review).
Hot accretion flow consists of optically thin, and geometrically thick disc
with very low mass accretion rate (e.g. \citealt{Narayan and Yi 1994, Narayan and Yi 1995};
\citealt{Blandford and Begelman 1999, Blandford and Begelman 2004}; \citealt{Yuan et al. 2012a, Yuan et al. 2012b};
\citealt{Mosallanezhad et al. 2014, Mosallanezhad et al. 2016}; \citealt{Zeraatgari and Abbassi 2015, Zeraatgari et al. 2018}). 
While, in cold accretion flow, 
the disc is optically thick with relatively high mass accretion rate. 

In terms of cold accretion flow, standard thin disc model is the 
first authentic model of black hole accretion disc (\citealt{Shakura and Sunyaev 1973}; 
\citealt{Novikov and Thorne 1973}, \citealt{Lynden-Bell and Pringle 1974}; \citealt{Pringle 1981}). 
In this model the heat generated by the viscosity locally radiates away from the
disc. Consequently, the disc temperature becomes far below the virial
temperature, i.e., $ 10^{4}-10^{7}\, K $. The criterion for mass accretion 
rate is the Eddington rate defined as 
$ \dot{M}_\mathrm{Edd} = L_\mathrm{Edd} / (\eta c^{2}) $, where $ L_\mathrm{Edd} $
is the Eddington luminosity, $ \eta $ is the radiative efficiency, and $ c $ is the speed of light.
The thin disc model can successfully be applied to many black hole systems 
when their mass accretion rate are slightly low, $ \dot{M} \la \dot{M}_\mathrm{Edd} $ 
(e.g., \citealt{Pringle 1981}; \citealt{Frank et al. 2002}; \citealt{Kato et al. 2008}; 
\citealt{Abramowicz and Fragile 2013}; \citealt{Blaes 2014};
\citealt{Koratkar and Blaes 1999}; \citealt{Remillard and McClintock 2006}; \citealt{McClintock et al. 2014}).

When the accretion rate is above the Eddington limit, advection becomes important, 
and the accretion flow can be described by the Super-Eddington (or supercritical) flow.
In this case, the radiative diffusion timescale, $ t_\mathrm{diff} $, can exceed the timescale for accretion, 
$ t_\mathrm{acc} $, as a consequence of the high mass accretion rate. 
Thus, the diffused photons cannot escape from the disc and accrete onto 
the black hole with gas particles. 
Note that in some 3D RMHD simulation of super-Eddington accretion (e.g., \citealt{Jiang et al. 2014}), 
they found that radiative transfer in the vertical direction is important thus photon trapping 
is not as strong as people previously thought.  

The HD and MHD numerical simulations of hot accretion flow have found 
that mass inflow rate decreases inward ( e.g., \citealt{Stone et al. 1999}; \citealt{Yuan et al. 2012b}). 
In this regards, various analytical works proposed to explain this result such as adiabatic 
inflow-outflow solution (ADIOS, \citealt{Blandford and Begelman 1999, Blandford and Begelman 2004}, 
\citealt{Begelman 2012}) 
and convection dominated accretion flow (CDAF, \citealt{Narayan et al. 2000}). 
Based on ADIOS model, mass loss in the outflow is the reason for 
the inward decrease of the mass accretion rate. Therefore, due to the 
presence of outflow the mass accretion rate is not a constant with 
radius and decreases towards the black hole. CDAF model also 
presented to explain the simulations is assumed to be convectively unstable.
However, recent numerical simulations have shown that MHD accretion flows 
are convectively stable (\citealt{Narayan et al. 2012}, \citealt{Yuan et al. 2012a}).

In the case of the supercritical accretion flow, the outflow/wind is unavoidable. Since
the accretion luminosity exceeds Eddington limit, the radiation force becomes much 
more greater than the gravity. Subsequently, at high latitudes, the gas particles can 
be accelerated by the radiation pressure and blown out from the system as 
multi-dimensional effects like jet/wind.
Some good candidates for supercritical accretion flows are ultra-luminous 
X-ray sources (ULXs), ultra-soft X-ray sources (ULSs), narrow-line Seyfert 1 galaxies (NLS1s), 
bright micro-quasars (see, e.g., \citealt{Wang and Zhou 1999}; \citealt{Boller 2000}; 
\citealt{Mineshige et al. 2000}; \citealt{Makishima et al. 2000}; \citealt{Miller et al. 2004}, 
\citealt{Done et al. 2007}; \citealt{Vierdayanti et al. 2010}; \citealt{Furst et al. 2016}; 
\citealt{Israel et al. 2017a, Israel et al. 2017b}; \citealt{Kaaret et al. 2017}; \citealt{Kosec et al. 2018}).

Several multi-dimensional/time-dependent radiation hydrodynamic (RHD), 
radiation magnetohydrodynamic (RMHD) and general relativistic-radiation 
magnetohydrodynamic (GR-RMHD) simulations have been performed to reveal the physical properties 
of the supercritical flows (\citealt{Eggum et al. 1987, Eggum et al. 1988}; 
\citealt{Okuda 2002}; \citealt{Okuda et al. 2005}; \citealt{Ohsuga et al. 2005}; 
\citealt{Ohsuga and Mineshige 2007}; \citealt{Ohsuga et al. 2009}; 
\citealt{Kawashima et al. 2009, Kawashima et al. 2012}; \citealt{Ohsuga and Mineshige 2011}; 
\citealt{Yang et al. 2014}; \citealt{Jiang et al. 2014}; \citealt{Sadowski et al. 2014, Sadowski et al. 2015}; 
\citealt{McKinney et al. 2014}; \citealt{Fragile et al. 2014}; 
\citealt{Takahashi et al. 2016};  \citealt{Kitaki et al. 2017}; \citealt{Kitaki et al. 2018}). 
The first one-dimensional analytical studies on super-Eddington accretion flow, i.e., 
the slim disc model have focused on the radial structure of the flow (\citealt{Begelman and Meier 1982}; \citealt{Abramowicz et al. 1988}; 
\citealt{Wang and Zhou 1999}; \citealt{Watarai and Fukue 1999}; \citealt{Mineshige et al. 2000}; 
\citealt{Watarai et al. 2000, Watarai et al. 2001}; \citealt{Watarai 2006}; \citealt{Fukue 2004}; \citealt{Gu and Lu 2007}). 
They used cylindrical coordinates ($ R, \phi, z $) and adopted 
$ H = c_\mathrm{s} / \Omega_\mathrm{K} $ for the disc height, 
where $ c_\mathrm{s} $ and $ \Omega_\mathrm{K} $ are the 
sound speed and the Keplerian velocity, respectively.
In the mentioned relation, based on hydrostatic equilibrium in 
the vertical direction, the disc height was considered constant. Although 
this approximation might be true for standard thin disc model, 
it is obviously inaccurate for supercritical disc where the disc is 
geometrically thick due to the high mass accretion rate. 
However, \citealt{Zeraatgari et al. 2016} solved the $ 1.5- $dimensional 
inflow-outflow equations of supercritical accretion flow
by assuming a power-law function for mass accretion rate, $ \dot{M}\propto r^s  $.
They found that $ s =1 $ due to the inclusion of the radiative cooling.
\citealt{Ohsuga et al. 2005} is one of the 
pioneer numerical simulation works which considered a relatively 
small angular momentum for the flow and obtained quasi-steady state solutions.
They found a small inflow region near the equatorial plane and very wide angle outflow region above the disc.

To reveal the complex two-dimensional structure 
and understanding the physical properties of the supercritical accretion flow,
\citealt{Gu 2012} adopted spherical polar coordinates 
and considered only $ T_{r\phi} $ component of the stress
tensor to mimic the angular momentum transfer by the magneto-rotational 
instability (MRI, \citealt{Balbus and Hawley 1998}). 
He assumed that the radiation pressure is much more stronger than the gas 
pressure, i.e., $ p_\mathrm{gas} / p_\mathrm{rad} \ll 1 $. He further assumed 
$ v_{\theta} = 0 $ which is obviously incorrect for the extremely high mass accretion rate.
By making use of the radial self-similar solutions, he showed that 
the polytropic relation adopted in previous analytical works was not suitable.
He found that even for marginally sub-Eddington accretion flow the energy 
advection was significant and the accretion disc was convectively stable.

In the present study, we revisit the inflow-outflow structure of the supercritical accretion flow by means of
radial self-similar solution. The main aim of this study is to relax the assumption of $ v_{\theta} = 0 $ 
compared with \citealt{Gu 2012} and consider very high mass accretion rate 
(see figure 1 of \citealt{Gu 2012} for more details). To do so, we adopt 
spherical coordinations and consider all components of the velocity, 
i.e., ($ v_{r}, v_{\theta}, v_{\phi} $) and also all components of 
the viscous stress tensor. We integrate the set of coupled 
RHD equations in the whole vertical angle, from rotational axis 
to the equatorial plane. Therefore, compare to the previous 
analytical works, we can clearly show two dimensional structure 
of the supercritical disc and address its physical properties when 
the disc is in the steady state.

The remainder of the manuscript is organized as follows. 
The basic equations and assumptions are described in section \ref{sec:equations}.
The self-similar solutions and boundary conditions are given in Section \ref{sec:self-similar}. In Section \ref{sec:results}, the numerical results are presented with detailed explanations. 
Finally, the summary and discussion are provided in Section \ref{sec:summary}.

\section{BASIC EQUATIONS AND ASSUMPTIONS} \label{sec:equations}

In this section, we describe the two-dimensional RHD equations of accretion 
flow around a non-rotating black hole in spherical coordinates ($ r, \theta, \phi $).
We neglect the self-gravity of the accretion disc. To avoid the relativistic effects,  
the Newtonian potential, $ \psi = -(GM) / r  $, is considered, where $ G $ is the 
gravitational constant and $ M $ is the black hole mass. The basic RHD equations 
of the accretion flow are written as follows,
\begin{equation} \label{eq:continuity}
	\frac{\partial \rho}{\partial t} + \bm{\nabla} \cdot \left( \rho \bm{v} \right)=0,
\end{equation}
\begin{equation} \label{eq:motion}
	\rho \left[ \frac{\partial \bm{v} }{\partial t} + \left( \bm{v} \cdot \bm{\nabla} \right) \bm{v} \right]
	= - \rho \bm{\nabla} \psi - \bm{\nabla} p_\mathrm{gas} +\bm{\nabla} \cdot \bm{\sigma} +  \frac{\chi}{c}\, \bm{F},
\end{equation}
\begin{equation} \label{eq:energy_gas}
	\frac{\partial e}{\partial t} + \bm{\nabla} \cdot \left( e \bm{v} \right) = 
	- p_\mathrm{gas} \bm{\nabla} \cdot \bm{v} - 4 \pi \kappa B + c \kappa E + \Phi_\mathrm{vis},
\end{equation}
\begin{equation} \label{eq:energy_rad}
	\frac{\partial E }{\partial t} + \bm{\nabla} \cdot \left(  E \bm{v} \right) = 
	- \bm{\nabla} \cdot \bm{F} - \bm{ \nabla v:P} + 4 \pi \kappa B - c \kappa E.
\end{equation}
In the above equations, $ \rho $ is the density, $ \bm{v} \left[= (v_{r},v_{\theta},v_{\phi})\right]$ is the velocity, 
$ p_\mathrm{gas} $ is the gas pressure, $ \bm{\sigma} $ is the viscous stress tensor, $ \chi  $ is the total opacity,
$ \bm{F} $ is the radiation flux, $ e $ is the internal energy density of the 
gas, $ E $ is the radiation energy density,  $ \bm{P} $ is the radiation pressure tensor, $ \kappa $ is
the absorption opacity, $ B $ is the blackbody intensity, and $ \Phi_\mathrm{vis} $ is the viscous 
dissipative function. The viscous stress tensor can be described as,
\begin{equation}\label{viscous_stress_tensor}
	\sigma_{ij} = \mu \left[ \left( \frac{\partial v_{j}}{\partial x_{i}} + \frac{\partial v_{i}}{\partial x_{j}} \right)  
	- \frac{2}{3} \left( \bm{\nabla} \cdot \bm{v} \right) \right]
\end{equation} 
where $ \mu (\equiv \nu \rho) $ is the dynamical viscosity coefficient which determines 
the magnitude of the stress and $ \nu $ is called the kinematic viscosity coefficient. 
We note that the bulk viscosity is neglected in this study. The dynamical viscosity coefficient 
is calculated with the usual $ \alpha $ prescription of the viscosity (\citealt{Shakura and Sunyaev 1973}),
\begin{equation}\label{mu_parameter}
	\mu =  \alpha \frac{p_\mathrm{gas} + \lambda E}{\Omega_{K} },
\end{equation}
where $ \alpha $ is the viscosity parameter, $ \lambda $ is the flux limiter, and 
$ \Omega_{K} \equiv (GM/r^{3})^{1/2}$ is the Keplerian angular momentum.
To calculate the radiation flux, $ \bm{F} $, we apply the flux-limited diffusion 
(FLD) approximation (\citealt{Levermore and Pomraning 1981}) as,
\begin{equation}\label{radiationflux}
  \bm{F} = -\frac{\lambda c}{\chi}\, \mathbf{\nabla} E, 
\end{equation}
where $ \lambda $ is flux limiter.
To avoid the complexity, the absorption opacity including both free-free absorption, 
$ \kappa_\mathrm{ff} $, and bound-free absorption, $ \kappa_\mathrm{bf} $, are 
neglected in the present study. The total opacity is then $ \chi = \rho \kappa_\mathrm{es} $, 
where $ \kappa_\mathrm{es} $ is the electron scattering opacity. The radiation pressure tensor 
is calculated in terms of energy density of the radiation as, 
\begin{equation} \label{radiation_pressure_tensor}
	 \mathbf{P} = \bm{f} E,  
\end{equation}
where $ \bm{f} $ is the Eddington tensor. In this study we focus on
optically thick region, where $ \lambda = 1/3 $  and $ \bm{f} = \bm{I} /3 $
(\citealt{Kato et al. 2008}). Thus, the Eddington approximation yields to,
\begin{equation}\label{initial_vp}
  \bm{P}_{ij} =
\begin{cases}
p_\mathrm{rad} = E/3 \quad \quad \quad &\text{if} \quad i = j,\\
0  &\text{if} \quad i \neq j.
\end{cases}
\end{equation}
By combining equation(\ref{eq:energy_gas}) and
equation(\ref{eq:energy_rad}), the total energy equation including gas and
radiation can be rewritten as,
\begin{multline} \label{eq:energy_total}
	\frac{\partial \left( e + E \right)}{\partial t} + \bm{\nabla} \cdot \left[ \left( e +  E \right) \bm{v} \right] = 
	- (p_\mathrm{gas} + p_\mathrm{rad}) \bm{\nabla} \cdot \bm{v}\\
	 - \bm{\nabla} \cdot \bm{F} + \Phi_\mathrm{vis},
\end{multline} 

where $ p_\mathrm{rad} $ is radiation pressure and based on our assumption it is much more stronger than the gas 
pressure, i.e., $ p_\mathrm{gas} / p_\mathrm{rad} \ll 1 $\footnote{In the future study we relax this 
assumption and work in a regime where gas pressure and also radiation pressure are comparable with each other.}. 
Thus, the gas pressure as well as internal energy density of the gas will be dropped  
in our equations. We consider steady-state and axisymmetric ($ \partial / \partial t = \partial / \partial \phi = 0) $
flow to solve equations (\ref{eq:continuity})-(\ref{eq:energy_rad}). 
The detailed form of partial differential equations (PDEs) are presented in Appendix \ref{appendix_A},
(see equations [\ref{continuity}]-[\ref{F_t}]). We outline our self-similar solutions and boundary conditions 
in the following section.

\section{Self-similar Solutions and Boundary Conditions} \label{sec:self-similar}

Many numerical simulations of accretion flow show that the radial profile of the density can be described as a power law function of $ r $ as $ \rho(r) \propto r^{-n} $. In terms of hot accretion flow, the global numerical simulations are consistent with the self-similar assumptions away from the boundaries (e.g., \citealt{Stone et al. 1999}; \citealt{Yuan et al. 2012a,Yuan et al. 2012b}; \citealt{Yuan et al. 2015}). For the case of super-Eddington accretion flow, recent numerical simulations also show that the radial profile of the density follows a power law form with $ n \approx 0.55 $ for a wide range of $ \alpha $ from $ \alpha = 0.005 $ to $ 0.1 $ (\citealt{Ohsuga et al. 2005}, \citealt{Yang et al. 2014}) \footnote{Note that \citealt{Kitaki et al. 2018} showed some deviations from the self-similar assumptions.}.
Therefore, in order to solve the equations ([\ref{continuity}]-[\ref{energy}]),
by numerical methods, we adopt self-similar solutions to remove 
the radial dependence of the variables. 

\subsection{Self-similar Solutions} \label{sub:self-similar}

By considering a fiducial radial distance, i.e., $ r_{0} $, 
the self-similar solutions can be written as a power-law 
form of $ r / r_{0} $. Thus, the physical variables of 
the flow can be written by the following radial scaling,

\begin{equation} \label{self_similar_vr}
	v_{r} \left(r,\theta \right) = v_{0} \left(\frac{r}{r_{0}} \right)^{-1/2} v_{r} (\theta),
\end{equation}

\begin{equation} \label{self_similar_vt}
	v_{\theta} \left(r,\theta \right) = v_{0} \left(\frac{r}{r_{0}} \right)^{-1/2} v_{\theta}(\theta),
\end{equation}

\begin{equation} \label{self_similar_vp}
	v_{\phi} \left(r,\theta \right) = v_{0} \left(\frac{r}{r_{0}} \right)^{-1/2} \Omega(\theta) \sin \theta,
\end{equation}

\begin{equation} \label{self_similar_rho}
	\rho \left(r,\theta \right) = \rho_{0} \left( \frac{r}{r_{0}} \right)^{-n} \rho(\theta),
\end{equation}

\begin{equation} \label{self_similar_p}
	p_\mathrm{rad} \left(r,\theta \right) = \rho_{0} v_{0}^{2} \left( \frac{r}{r_{0}} \right)^{-n-1} p(\theta),
\end{equation}
where $ v_{0}  = (GM/r_{0} )^{1/2}$ and $ \rho_{0} $ are 
considered to be Keplerian velocity and density at $ r_{0} $, 
respectively. By substituting above self-similar solutions into 
equations [\ref{continuity}]-[\ref{energy}],
the radial dependency will be removed only if $ n = 1/2 $. 
This is mainly due to the inclusion of the radiative cooling 
in the energy equation, i.e., $ \bm{\nabla} \cdot \bm{F} $. 
Consequently, the radial profile of the accretion rate can be well 
described by $ \dot{M} \propto r  $ which is fully consistent 
with hyperaccreting ADIOS model of \citealt{Begelman 2012}.
This result is again consistent with the radial dependency of 
the density found in this present study. 

Substituting equations (\ref{self_similar_vr})-(\ref{self_similar_p})
into equations ([\ref{continuity}]-[\ref{energy}], we can reduce them 
to ordinary differential equations (ODEs) given in Appendix 
(\ref{appendix_B}). Equations (\ref{ode_continuity})-(\ref{ode_energy}) 
describe the variations of the five physical quantities, $ v_{r} (\theta) $, 
$ v_{\theta}(\theta) $, $ v_{\phi}(\theta) $, $ \rho(\theta) $, and 
$ p(\theta) $ in the vertical direction. 

\begin{figure*}[ht!]
\includegraphics[width=\textwidth]{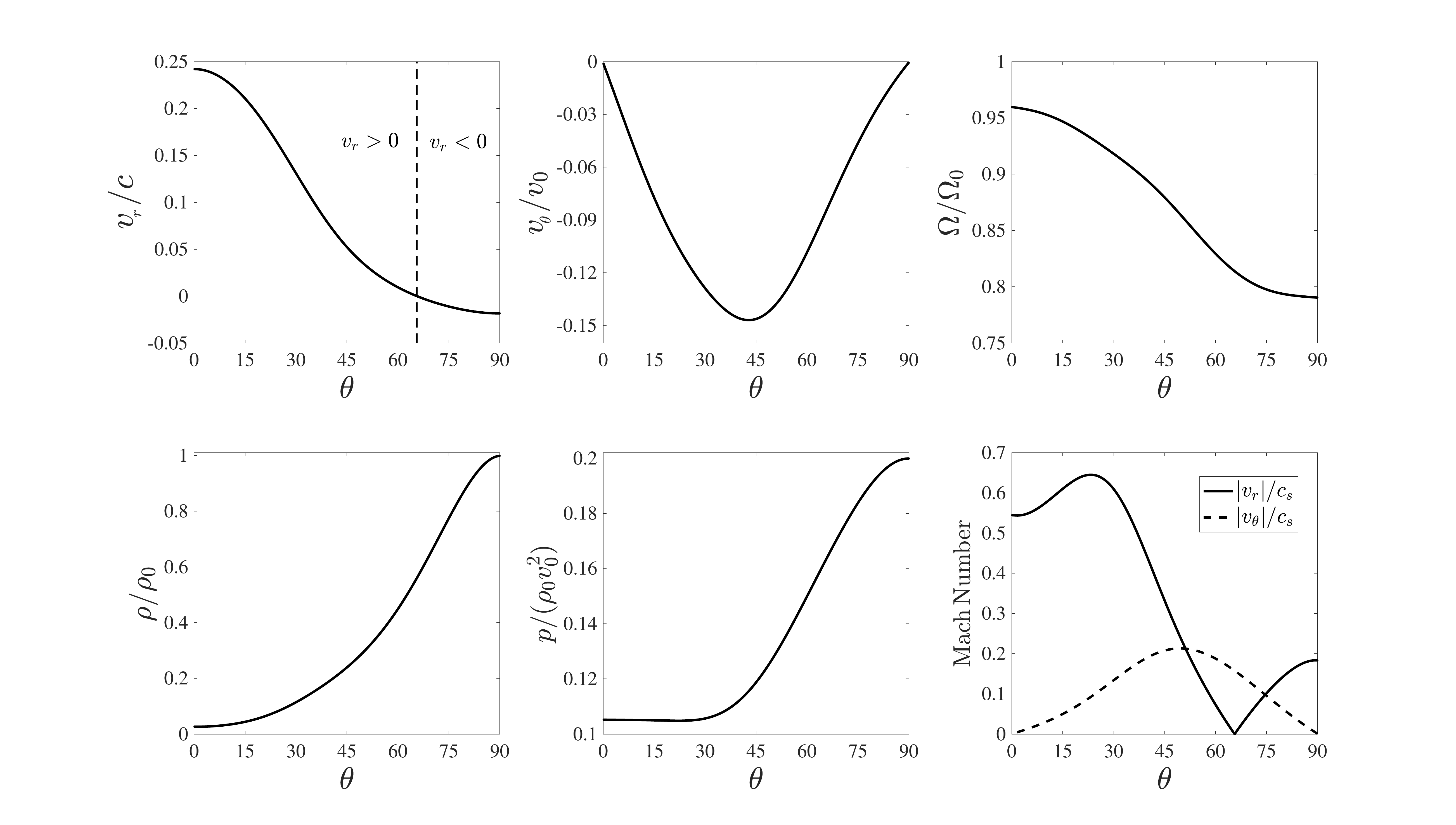} 
\caption{Angular profile of the physical variables at $ r = 10 r_{s} $. 
\textit{Top-left panel:} radial velocity in the unit of light speed, $ c $.
Dashed line shows the location of $ v_{r} = 0 $ that is about $ 65^{\circ} $. 
\textit{Top-middle panel:} latitudinal velocity in the unit of Keplerian 
velocity, $ v_{0} $. \textit{Top-right panel:} Angular velocity in the unit 
of Keplerian angular velocity, $ \Omega_{0} $.  \textit{Bottom-left panel:}  
density in the unit of density of mid-plane at $ r_{0} $, i.e., $ \rho_{0} $. 
\textit{Bottom-middle panel:} radiation pressure in the unit of $ \rho_{0} v_{0}^2 $. 
\textit{Bottom-right panel:} Mach numbers.
\label{physical_variables}}
\end{figure*}

\begin{figure*}[ht!]
\includegraphics[width=0.5\textwidth]{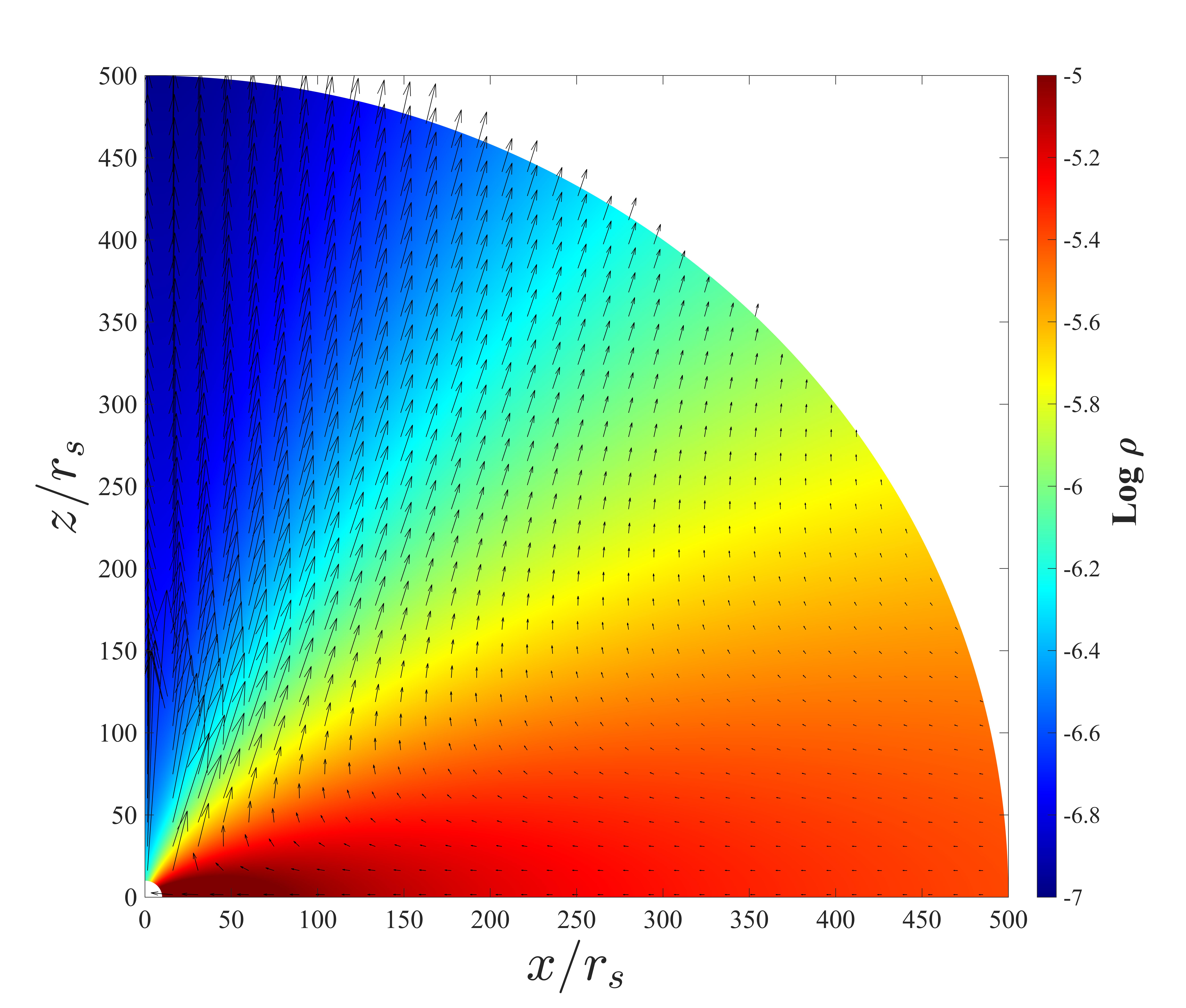} 
\includegraphics[width=0.5\textwidth]{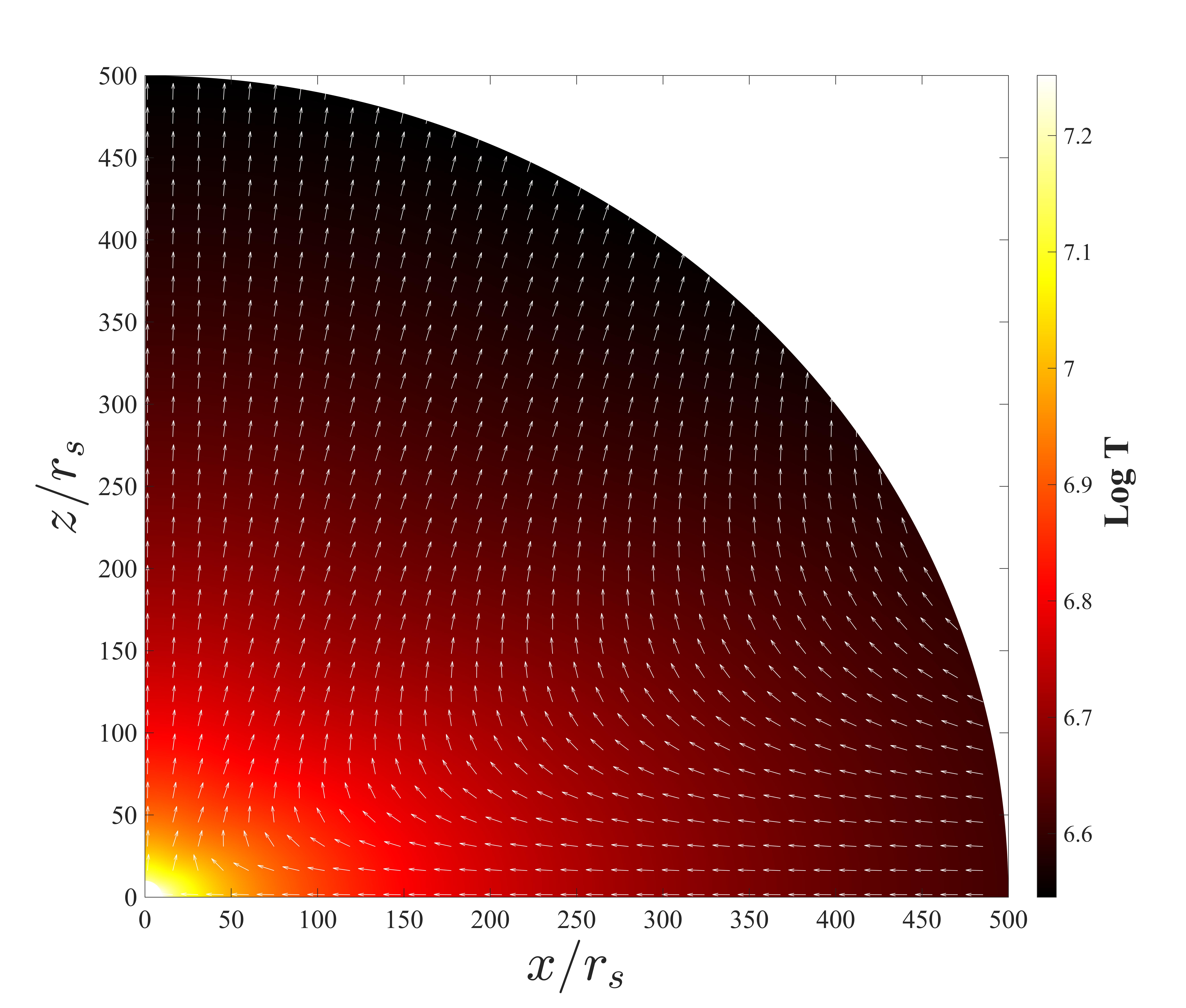}
\caption{Two-dimensional distribution of the density (left panel) 
and the temperature (right panel) based on the self-similar solutions. 
Both panels overlaid with the poloidal velocity. In the left panel,
the poloidal velocity is normalized by $ 0.1 c $ to denote the strength 
of the outflow while in the right panel, the poloidal velocity is normalized with 
its absolute value to denote the direction of the vectors.
 \label{density_temperature}}
\end{figure*}

\subsection{Boundary Conditions}

Following \citealt{Narayan and Yi 1995}, we assume that all flow variables 
are even symmetric, continuous, and differentiable at the equatorial 
plane, $ \theta = \pi /2 $, and the rotation axis, $ \theta = 0 $. The main 
difference here compare to the previous works is that we include 
the latitudinal component of the velocity, $ v_{\theta} $, in our 
equations and consider it to be zero at both the equatorial plane 
and the rotation axis. Therefore, we apply the following boundary 
conditions at $  \theta = \pi /2 $ and $ \theta = 0 $:
\begin{equation}
v_{\theta} = \frac{\mathrm{d} \rho}{\mathrm{d} \theta} = \frac{\mathrm{d} p}{\mathrm{d} \theta} = 
\frac{\mathrm{d} v_{r}}{\mathrm{d} \theta} = \frac{\mathrm{d} \Omega}{\mathrm{d} \theta} = 0
\end{equation}
 
To solve the set of ODEs, (see equations [\ref{ode_continuity}]-[\ref{ode_energy}] 
in Appendix \ref{appendix_B}), we need to set fiducial radial distance, 
$ r_{0} $ and the density there, $ \rho_{0} $, which are defined in equations 
(\ref{self_similar_vr})-(\ref{self_similar_p}), (see the constant term, $ \tau^{-1} (c/v_{0}) $, 
in the right-hand side of the equation [\ref{ode_energy}] as well). 
Numerical simulations of accretion flow show that the radial velocity 
increases inward very rapidly because of the strong gravity near the black 
hole (\citealt{Ohsuga et al. 2005}, \citealt{Yuan et al. 2012a, Yuan et al. 2012b}). 
To avoid shock occurring by supersonic inflow near the central region, 
which is a source of deviation from the self-similar assumptions, we neglect 
the region within $ 10 r_\mathrm{s} $, where $ r_\mathrm{s} = 2 G M / c^2 $ is 
the Schwarzschild radius (see e.g., \citealt{Kawashima et al. 2012}, \citealt{Jiang et al. 2014}).
Due to the assumption of high mass accretion rate, 
we expect strong radiation produced at the innermost region interacts 
with gas particles at the region $ r \gtrsim 10 r_\mathrm{s} $ and strong 
outflow is driven there. To show the two dimensional inflow-outflow 
structure of the flow, we consider all components of the velocity including 
$ v_{\theta} $ in our equations. Throughout present study, we set the 
inner and outer radial range of the domain as $ r_\mathrm{min} = 10 r_\mathrm{s} $
and $ r_\mathrm{max} = 500 r_\mathrm{s} $, respectively. Thus, the 
assumption of Newtonian potential is safely valid in this range.  We set 
$ r_{0} = 10 r_\mathrm{s} $ and for the initial density, $ \rho_{0} $, we 
calculate the mass inflow rate at the outer radial boundary.
Following numerical simulations of supercritical accretion flow, 
the dimensionless mass inflow rate defines as
(see \citealt{Ohsuga et al. 2005}, \citealt{Yang et al. 2014})
\begin{equation} \label{mdot}
	\dot{m} = - \frac{c^2}{L_\mathrm{Edd}} \int_{0}^{\pi}  2 \pi r^{2} \rho\, \mathrm{min} 
	\left(v_{r},0 \right) \, \sin \theta\,  d \theta. 
\end{equation}

We set $ \dot{m} = 1300 $ at the outer radial boundary, i.e., 
$ r_\mathrm{max} = 500 r_\mathrm{s} $ throughout this paper. 
Based on self-similar solutions adopted here, the mass inflow 
rate in this study is not radially constant and decreases inward 
as $ \dot{m} \propto r $. We use iteration method to find the 
value of $ \rho_{0} $ by solving equation (\ref{mdot}) at the outer 
radial boundary.

\section{Numerical results} \label{sec:results}

We solve the ordinary differential equations 
[\ref{ode_continuity}]-[\ref{ode_energy}] by integrating
from the equatorial plane ($ \theta = \pi/2 $) to the rotation 
axis ($ \theta = 0 $). We adopt the values of $ \alpha = 0.1 $, 
$ M = 10 M_{\odot} $, and $ \kappa_\mathrm{es} = 0.34 $ with 
reference to the numerical simulations of \citealt{Ohsuga et al. 2005} 
and \citealt{Yang et al. 2014}. The main difference here compare to 
those numerical simulations is that we consider all components 
of viscous stress tensor.  As it is expressed in previous section, 
the radial range of our calculation is $ 10 r_\mathrm{s}  \leq r \leq 500 r_\mathrm{s} $. 
We implement relaxation method to solve the set of equations 
along the vertical direction. The grid in $ \theta $ direction is divided 
into 2000 equally spaced points and iteration technique is 
used to find the value of $ \rho_{0} $ with $ \dot{m} = 1300 $ 
at $ r_\mathrm{max} $. For this constant mass accretion rate, 
we obtain $ \rho_{0} = 2.89 \times 10^{-5} \mathrm{gr/cm^3}$.  
We can reasonably treat this value as a boundary condition. 
The global properties of the solutions we obtain in this way agree 
well with those presented in numerical simulations of \citealt{Ohsuga et al. 2005} 
and \citealt{Yang et al. 2014}. In the following subsections, we explain in 
details the flow properties based upon our solutions. 

\subsection{Inflow-outflow structure of the solutions}

Figure \ref{physical_variables} presents angular profiles of 
physical quantities at $ r=10 r_\mathrm{s} $. We can see, 
the density and the radiation pressure decrease rapidly from 
the equatorial plane to the rotation axis. Since we are interested 
in studying the case where the radiation pressure is much more stronger 
than the gas pressure, i.e., $  p_\mathrm{gas} / p_\mathrm{rad} \ll 1 $, 
this pressure represents the total pressure of the flow. As it 
is shown in the top-left panel of Figure \ref{physical_variables}, 
at the region close to the equatorial plane, $ \theta > 65^{\circ} $, 
the radial velocity is negative and the gas particles move 
toward the central black hole. While, in the region, 
$ \theta < 65^{\circ} $, the sign of the radial velocity 
changes and becomes positive. Furthermore, the top-middle 
panel shows the variation of $ v_{\theta} $ along 
the vertical direction. From this plot, it is seen that $ v_{\theta} $ 
has a negative value in all angles, and is zero at both the 
equatorial plane and the rotation axis due to the boundary conditions. 
The minimum value of $ v_{\theta} $ is also located around 
$ \theta \sim 43^{\circ} $. Moreover, the bottom-right panel of 
Figure \ref{physical_variables} represents the Mach numbers.
We plot this figure to check the existence of the sonic point at 
the high-latitude region. As we can see, there does not exist 
any sonic points in this region. The Mach number $ |v_{r}| / c_{s} $ 
decreases from the equatorial plane to about $ 65^{\circ} $, where 
the radial velocity is zero. Then, it increases until $ \theta = 30^{\circ} $ 
and again decreases rapidly to the rotation axis.  This behavior 
can be explained by the isothermal sound speed, $ c_\mathrm{s} = p / \rho $. 
The profile of the density and the total pressure in Figure \ref{physical_variables} 
show that the density decreases rapidly from the equator to the pole 
while the pressure is almost constant at the range $ 0 < \theta < 30^{\circ} $. 
Therefore, the Mach number declines in this range. Also, $ |v_{\theta}|/c_\mathrm{s} $ 
has a maximum value at $ \theta = 43^{\circ} $ and becomes 
null at both axes due to the boundary conditions there, i.e, 
$ v_{\theta} = 0 $. For both lines, the Mach numbers are 
less than unity which clearly shows that there is no critical point at high latitudes.

In the left panel of Figure \ref{density_temperature}, we plot 
two-dimensional distribution of the density overlaid with the 
stream lines of the flow. The results are shown for $ \dot{m} = 1300 $ 
and the poloidal velocity is normalized by $ 0.1c $. It can be 
seen, the density tends to be larger around mid-plane than 
that around the polar axis. Moreover, the stream lines are 
directed toward the black hole at low latitudes with small magnitudes. 
At high latitudes, they are pointed outward and become outflow. 
For high accretion rate, as in our present study, the outflow can 
be driven by the radiation pressure produced at inner radii. 
This strong radiation interacts with the gas particles and can push 
them away as high velocity outflow. This figure clearly shows 
the high velocity field region at high latitudes at $ r=10 r_\mathrm{s} $.  
As shown in top-left and top-middle panels of figure \ref{physical_variables},
the poloidal velocity of outflow reaches to about $ \sim 0.25 c $. 
These results are fully consistent with the results obtained 
in \citealt{Yang et al. 2014} (see Fig. 1) and \citealt{Ohsuga et al. 2005}, 
where the inflow is presented in low latitude around the equatorial 
plane of the disc while outflow is presented at high latitude regions. 

The two-dimensional profile of the radiation temperature, 
$ T = ( E/ a)^{1/4} $, is plotted in the right panel of 
Figure \ref{density_temperature}, (where, $ a = 7.5646 \times 
10^{-15}\, \mathrm{erg}\, \mathrm{cm}^{-3}\, \mathrm{K}^{-4} $ 
is the radiation constant). The logarithm of the temperature is 
overlaid with the poloidal velocity normalized with its absolute 
value denotes the direction of the vectors. With the assumption 
of $ p_\mathrm{gas}/p_\mathrm{rad} \ll 1 $, the gas and the radiation 
are in equilibrium, so their temperatures are almost equal.
Due to the heating of the gas by the viscous dissipation, 
the temperature is relatively higher in the inner region than 
that at the outer radii. Therefore, this produced energy can be 
effectively converted into the radiation energy. This figure 
somehow represents the distribution of the radiation internal 
energy density. In addition, at large radii with high latitudes, 
we can see the temperature is lower than one at low latitudes with
the same radii. This is the consequence of the radiative cooling by the outflow.

\begin{figure*}[ht!]
\includegraphics[width=0.5\textwidth]{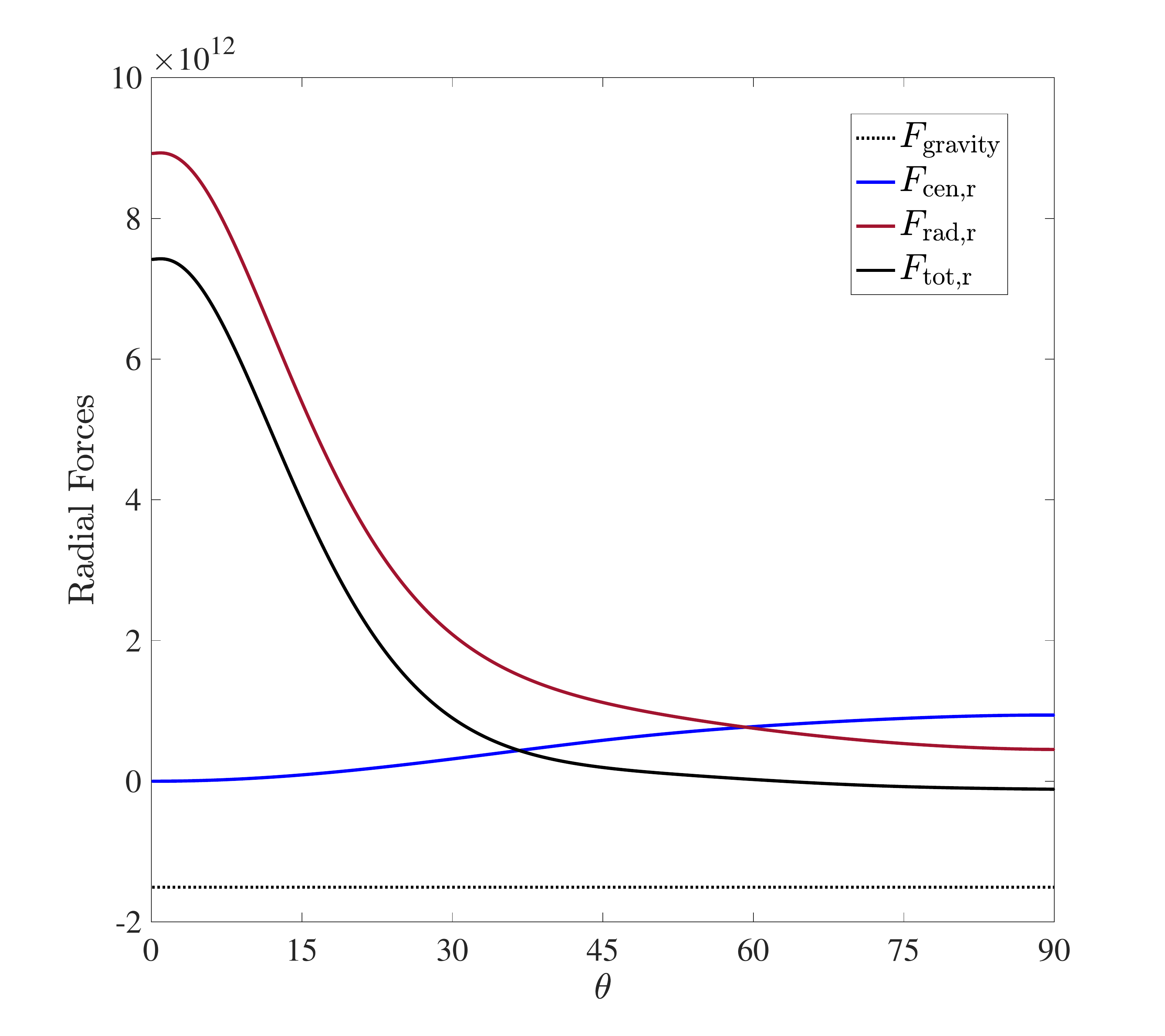}
\includegraphics[width=0.5\textwidth]{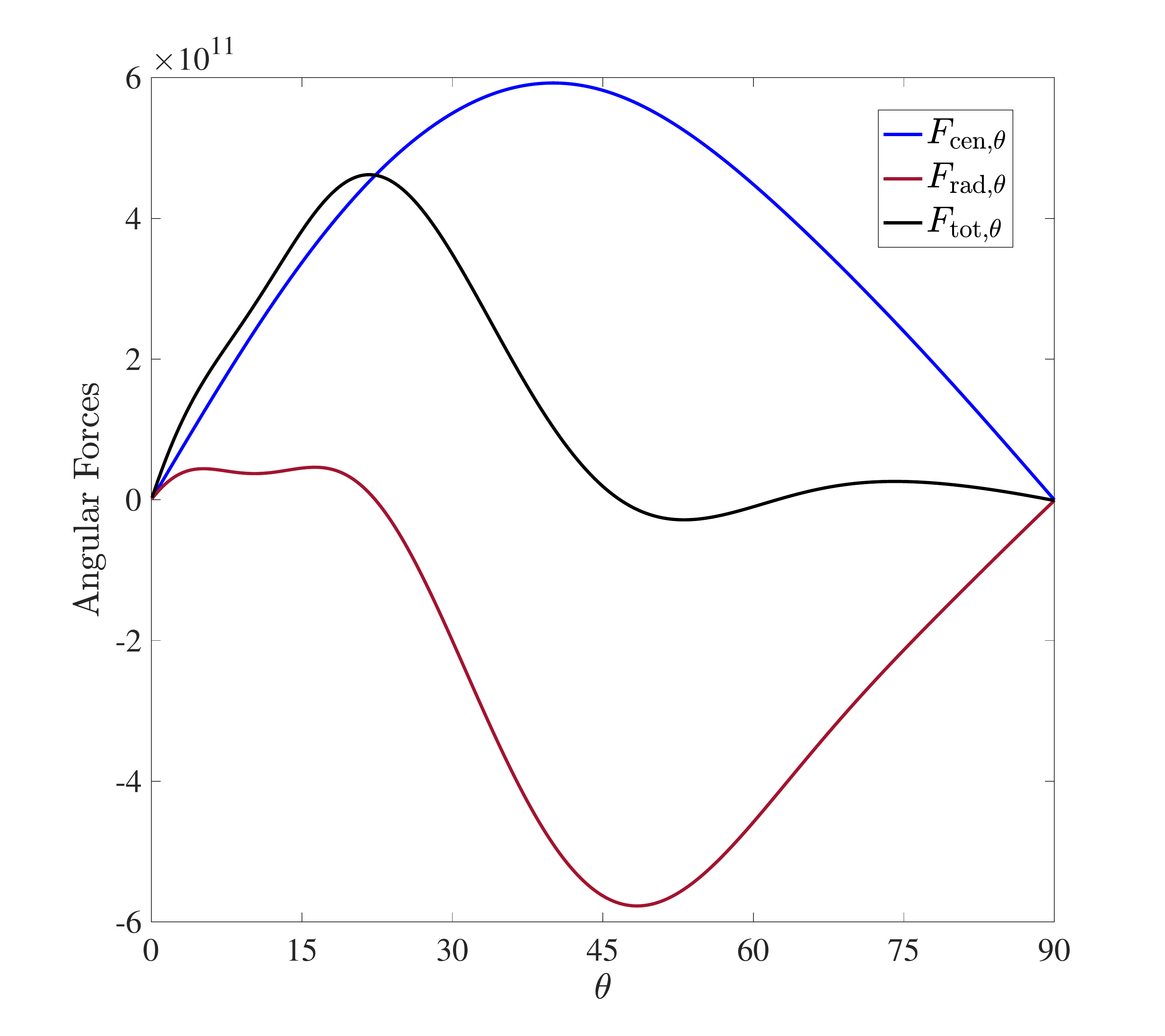} 
\caption{Angular distribution of the radial forces (left panel), and
angular forces (right panel) per unit mass at $ r = 10 r_\mathrm{s} $.  
The forces include gravity (black dotted line), centrifugal force 
(blue solid line), radiation force (red solid line), and their sum 
(black solid line). \label{radial_angular_forces}}
\end{figure*}

\begin{figure*}[ht!]
\begin{center}
\includegraphics[width=0.6 \textwidth]{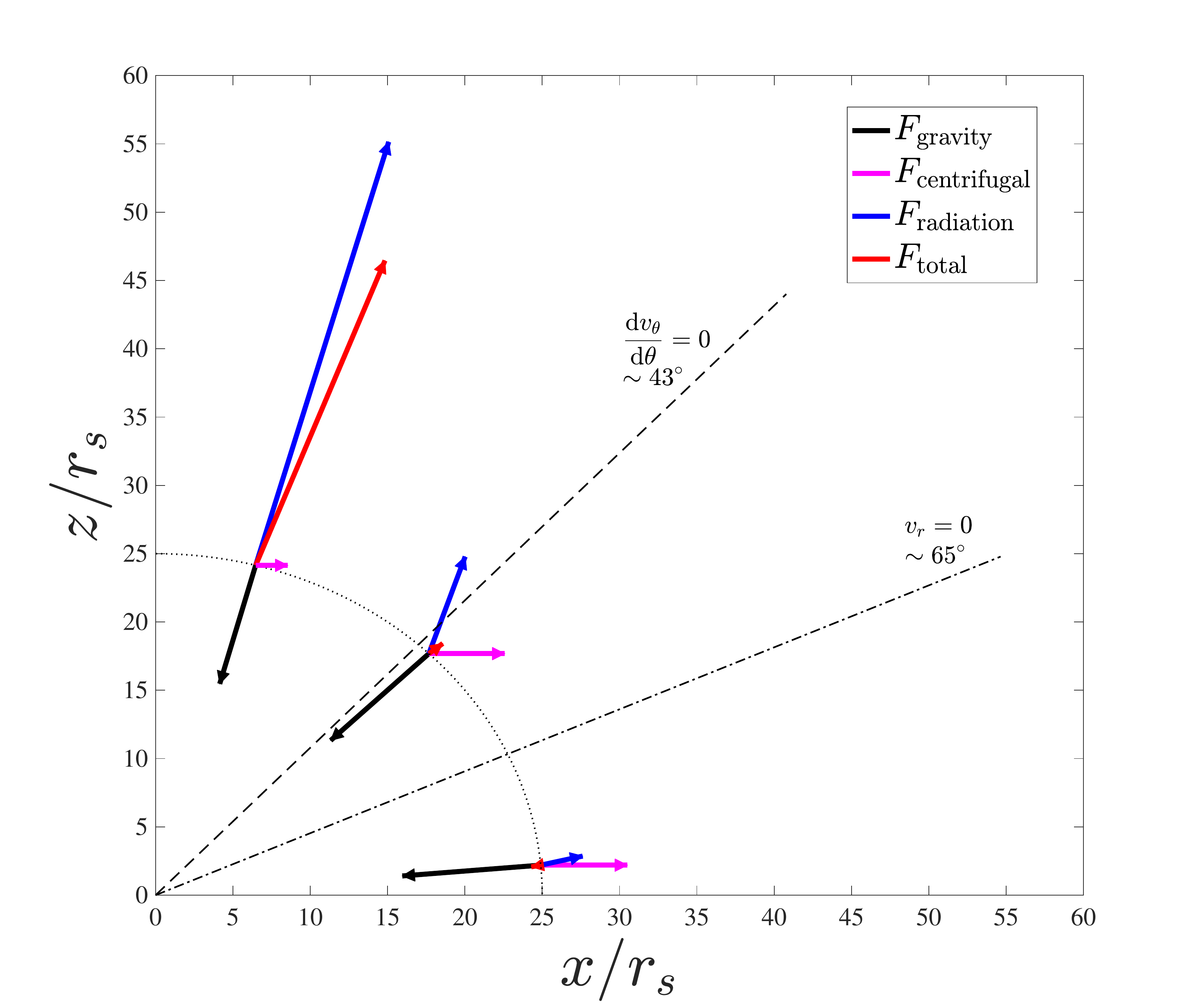} 
\caption{Force analysis at inflow/outflow region to show the driving 
mechanism of the outflow at $ r = 25 r_\mathrm{s} $. The length 
of the arrows schematically denotes the magnitude of the forces 
while the direction for the direction of forces. The forces include gravity, 
centrifugal, radiation, and the sum of them. The dash-dotted line shows the
location of $ v_{r} = 0 $, the dashed line represents the location of 
$ \frac{\mathrm{d} v_{\theta}}{\mathrm{d} \theta} = 0 $, and dotted line
shows the radius where the forces calculated.
\label{force_analysis}}
\end{center}
\end{figure*}

\subsection{Physical properties of the outflow}

In this subsection we calculate the physical properties of the 
outflow based on our self-similar solutions. In the top-right panel 
of Figure \ref{physical_variables} the angular velocity of the 
accretion flow is plotted. It is seen that the angular velocity 
increases from the equatorial plane to the rotation axis and 
becomes almost Keplerian at high latitudes which is 
consistent with \citealt{Yuan et al. 2012a} numerical simulation results. This indicates 
that the outflow can transport angular momentum outward 
from the disc. To check this issue in more detail, we evaluate 
the mass flux weighted value of the inflow and the outflow 
quantities as,
\begin{equation} \label{mass_flux_inflow}
	Q_\mathrm{in}(r) = \frac{4 \pi r^{2} \int^{\pi/2}_{0} \rho Q \, \mathrm{min}(v_{r},0) \sin \theta d\theta}
	{4 \pi r^{2} \int^{\pi/2}_{0} \rho \, \mathrm{min}(v_{r},0) \sin \theta d\theta},
\end{equation}
\begin{equation} \label{mass_flux_outflow}
	Q_\mathrm{out}(r) = \frac{4 \pi r^{2} \int^{\pi/2}_{0} \rho Q \, \mathrm{max}(v_{r},0) \sin \theta d\theta}
	{4 \pi r^{2} \int^{\pi/2}_{0} \rho \, \mathrm{max}(v_{r},0) \sin \theta d\theta},
\end{equation}
where, $ Q $ represents the physical quantities. The mass flux weighted angular momentum of 
the inflow and the outflow are found as,
\begin{equation}
	L_\mathrm{in} = 0.79 \, L_{K}, 
\end{equation}
and
\begin{equation}
	L_\mathrm{out} = 0.89 \, L_{K}, 
\end{equation}
where $ L_{K} $ is the Keplerian angular momentum. 
This result clearly shows that the angular momentum 
of the flow can be transferred by the outflow which is 
fully consistent with the numerical simulation results of 
supercritical accretion discs (see \citealt{Ohsuga et al. 2005} and 
\citealt{Yang et al. 2014}). We also define the Bernoulli parameter as,

\begin{equation}
	Be(r) = \frac{v^{2}}{2} + \frac{\gamma p}{\left(\gamma - 1 \right)\rho} - \frac{GM}{r}, 
\end{equation}
where, $ \gamma = 4/3 $ is the specific heat ratio. 
The Bernoulli parameter is the sum of
the kinetic energy, the enthalpy, and the gravitational 
energy of the accreting gas. The mass flux 
weighted Bernoulli parameter of the outflow is obtained as, 
\begin{equation}
	Be = 2.67 \, v_{\!K}^{2}.
\end{equation}

The value of the Bernoulli parameter is positive which shows the outflow has 
enough energy to overcome the gravity of the central back hole and escape to
the infinity. The mass flux weighted poloidal velocity of the outflow, i.e., 
$ v_\mathrm{pol}^{2}  = v_{r}^{2} + v_{\theta}^{2} $, is also calculated as,

\begin{equation}
	v_\mathrm{pol} = 0.44  \, v_{\! K}.
\end{equation}

Here, we conclude that the radiation driven outflow has enough energy 
and power to interact with its surrounding, overcome the gravitational potential and 
escape to the infinity. 

\subsection{Analysis of forces driving the outflow}

In order to understand which force can drive outflow from our system,
we plot in Figure \ref{radial_angular_forces} the angular distribution of
the radial forces (left panel) and the angular forces (right panel) at 
$ r = 10 r_\mathrm{s} $. We can see from the left panel of this figure 
that the angular profile of the total force has similar behavior with the 
angular profile of the radial velocity shown in Figure \ref{physical_variables}. 
In addition, left panel of Figure \ref{radial_angular_forces} shows that 
within $ 60^{\circ} < \theta < 90^{\circ} $ the radial component of the
centrifugal force is greater than the radial component of 
the radiation force and can effectively counteracts the gravitational 
force in this range. However, within $ 0^{\circ} < \theta < 60^{\circ} $, 
the radial component of the centrifugal force decreases rapidly and
becomes null at the rotation axis. In contrast, the radial component of the radiation 
force is the dominant force near the rotation axis and we can see 
there exists very strong outflow near this axis. Therefore, it can be concluded 
that the radial radiation force is the dominant force and plays the important 
role to drive outflow at high latitudes. 
It is interesting here to compare this result with the case of hot accretion flows
(\citealt{Yuan et al. 2015}), in which radiation can be neglected. Although the dominant 
driving force is different in the two cases, outflow is always present and even 
their properties are similar. 

It is seen from left panel of Figure \ref{radial_angular_forces}, the angular 
component of the total force is very small at the region near the equatorial plane, 
i.e., $ 50^{\circ} < \theta < 90^{\circ} $, which clearly shows that the flows 
are in the force equilibrium in the inflow region. This is mainly because from 
this panel we can see that the centrifugal force balances with the radiation 
force in the vertical direction. The angular distribution of both the radiation force 
and the centrifugal force become zero at both axes due to the boundary conditions. 

To have a better understanding of the magnitude of the forces in different 
$\theta$ angels and study the driving mechanisms of the outflow, we calculate 
the forces at different regions. Figure \ref{force_analysis} shows the result at 
$ r = 25 r_\mathrm{s} $ in the unit of gravity. We can see from this figure that 
at the inflow region, $ \theta = 85^{\circ} $, the dominant force is the gravity
so, the flow moves toward the central black hole. In the 
intermediate region, $ \theta  = 45^{\circ} $, the driving forces are the 
centrifugal and the radiation forces. As it is seen, the strength of these forces 
are comparable means both of them can drive 
outflow in this region, but not so strong. Instead, at very high latitudes, 
$ \theta = 15^{\circ} $, the radiation force can efficiently offset the gravity
and play a noticeable role in driving the outflow whereas the centrifugal 
force is negligible and does not have any efficient contribution to the total force. 
These results are again fully consistent with those found in \citealt{Yang et al. 2014} 
and \citealt{Ohsuga et al. 2005}. 

\begin{figure}[ht!]
\begin{center}
\includegraphics[width=0.5 \textwidth]{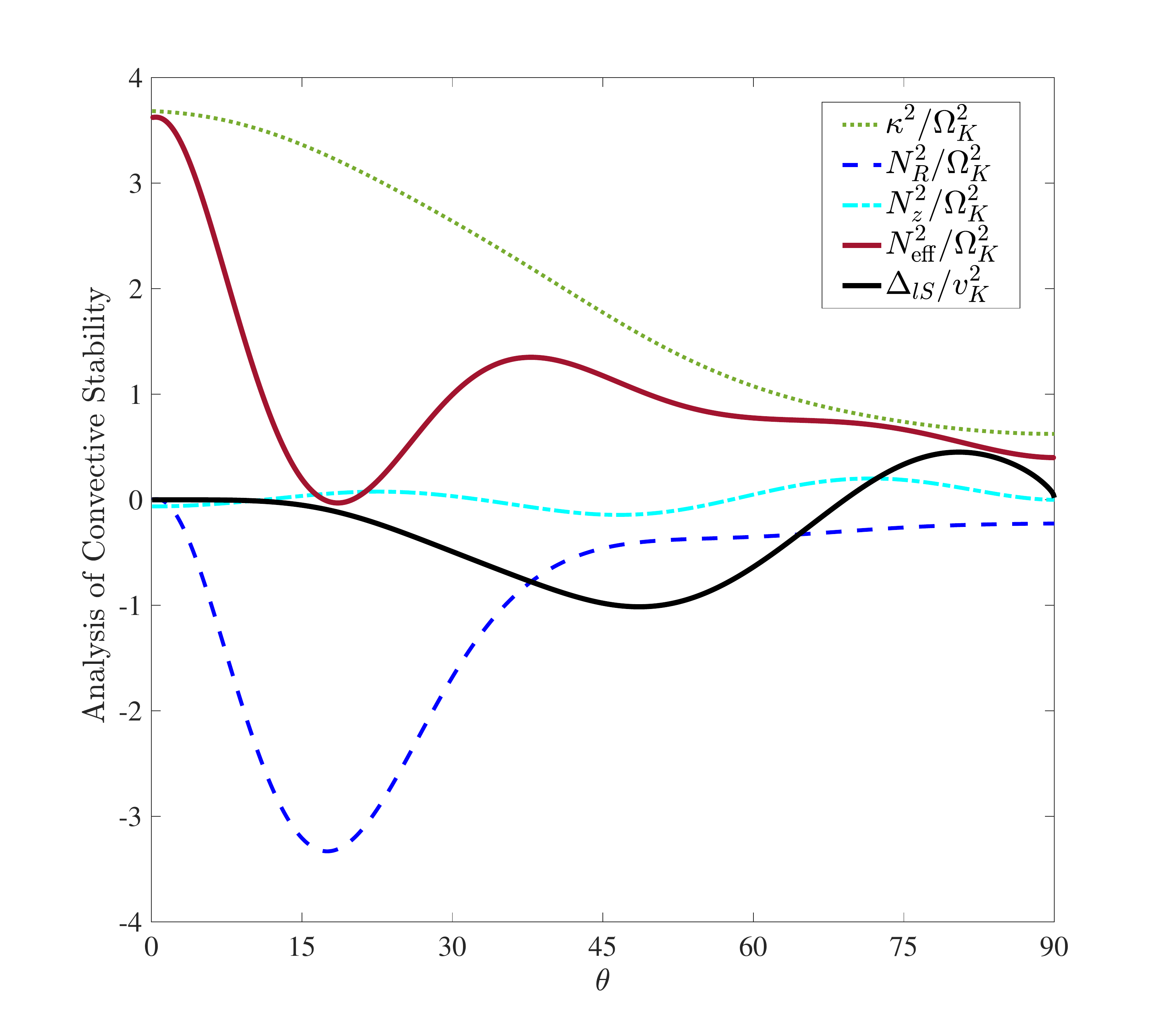} 
\caption{Angular variations of $ \kappa^{2} $ ( green dotted line), 
$ N_{R}^{2} $ (blue dashed line), $ N_{z}^{2} $ (cyan dash-dotted line), 
$ N^{2}_\mathrm{eff} $ (red solid line), and $\Delta_{lS} $ 
(black solid line). The quantities $ \kappa^{2} $, $ N_{R}^{2} $ , 
$ N_{z}^{2} $ , and $ N^{2}_\mathrm{eff} $ are normalized by 
$ \Omega_{K}^{2} $, and $\Delta_{lS} $  is normalized by $ v_{\!{K}}^{2} $. 
\label{convective}}
\end{center}
\end{figure}

\subsection{Convective stability}

In term of supercritical accretion flow, due to the large scattering optical depth, 
the photons produced from innermost part of the disc are trapped and cannot 
efficiently escape from the disc. Furthermore, the specific entropy is dominated 
mostly by the radiation photons since the radiation pressure is much more 
important than the gas pressure here. Several analytical and simulation 
works have been done to study the convective stability of supercritical accretion flows. 
For instance, based on the local energy balance, \citealt{Sadowski et al. 2009} and \citealt{Sadowski et al. 2011} 
found that the disc is convectively unstable. However, \citealt{Gu 2012} used self-similar 
solutions and concluded that radiation pressure-supported disc is always convectively stable. 
This discrepancy between these works might be related to the different definition of
vertical structure of the disc. 
In terms of numerical simulation, \citealt{Yang et al. 2014} studied this issue
based on their simulation data for large and small values of viscosity parameter. 
They found that for large value of viscosity parameter, $ \alpha = 0.1 $, about half 
of the computational domain was convectively unstable, while for $ \alpha = 0.005 $,
the fraction of the unstable region was much less. They concluded that radiation 
plays an effective role to stabilize the convection and can directly transport energy.
 
In this subsection, we revisit this problem and analyze the convective stability of 
supercritical accretion flow, in the absence of magnetic field, based on our self-similar 
solutions. Thus, we use the well-known Solberg-H\o iland criterions in cylindrical 
coordinates $ (R, \phi, z) $ as follows,

\begin{equation} \label{Hoiland_1}
	\frac{1}{R^{3}} \frac{\partial l^{2}}{\partial R} - \frac{1}{C_{P} \rho} 
	\bm{\nabla} P \cdot \bm{\nabla} S > 0,
\end{equation}

\begin{equation} \label{Hoiland_2}
	- \frac{\partial P}{\partial z} \left( \frac{\partial l^{2}}{\partial R} 
	\frac{\partial S}{\partial z} -  \frac{\partial l^{2}}{\partial z} \frac{\partial S}{\partial R} \right) > 0,
\end{equation}
where $ l (= r \sin \theta v_{\phi}) $ is the specific angular momentum per unit mass, 
$ C_{P} $ is the specific heat at constant pressure, $ P $ is the total pressure, and 
$ S $ is the entropy defined as,

\begin{equation} \label{entropy}
	dS \propto d \, \ln \left(\frac{P}{\rho^{\gamma}} \right).
\end{equation}
As we expressed in the previous section, we ignored the gas pressure in this study,
therefore the total pressure is equal to the radiation pressure, i.e., $ P = E / 3 $.
The first Solberg-H\o iland criterion can be simplified as

\begin{equation} \label{Hiland_1_simplified}
	N_\mathrm{eff} = \kappa^{2} + N_{R}^{2} + N_{z}^{2} > 0,
\end{equation}
with
\begin{equation} \label{kappa}
	\kappa^{2} = \frac{1}{R^{3}} \frac{\partial l^{2}}{\partial R}, 
\end{equation}
\begin{equation} \label{N2_R}
	N^{2}_{R} = - \frac{1}{\gamma \rho} \frac{\partial P}{\partial R} 
	\frac{\partial}{\partial R} \ln \left( \frac{P}{\rho^{\gamma}} \right),
\end{equation}
\begin{equation} \label{N2_z}
	N^{2}_{z} = - \frac{1}{\gamma \rho} \frac{\partial P}{\partial z} 
	\frac{\partial}{\partial z} \ln \left( \frac{P}{\rho^{\gamma}} \right).
\end{equation}

Here, $ N_\mathrm{eff} $ is the effective frequency, $ \kappa $ is the epicyclic frequency, and
$ N^{2}_{R}$ and $ N^{2}_{z} $ are defined as the $ R $ and $ z $ component of the Brunt-V\"{a}is\"{a}l\"{a} 
frequency, respectively. Based on our self-similar solutions, $ \partial P / \partial z $ is always negative. 
Therefore, the second Solberg-H\o iland criterion can be reduced as,

\begin{equation} \label{Hiland_2_simplified}
	\Delta_{lS} \equiv \frac{\partial l^{2}}{\partial R} \frac{\partial}{\partial z} \ln \left( \frac{P}{\rho^{\gamma}} \right)  -  \frac{\partial l^{2}}{\partial z} \frac{\partial}{\partial R}  \ln \left( \frac{P}{\rho^{\gamma}} \right)  > 0.
\end{equation}

To find the angular dependency of the two Solberg-H\o iland criterions in spherical coordinates, 
we apply the following transformations:

\begin{equation} \label{transformation_R}
	\frac{\partial}{\partial R} = \sin \theta \frac{\partial}{\partial r} + \frac{\cos \theta}{r} \frac{\partial}{\partial \theta}
\end{equation}
\begin{equation} \label{transformation_z}
	\frac{\partial}{\partial z} = \cos \theta \frac{\partial}{\partial r} - \frac{\sin \theta}{r} \frac{\partial}{\partial \theta}.
\end{equation}

 The results are shown in Figure \ref{convective}. In this figure we plot the angular 
variations of  $ \kappa^{2} $, $ N^{2}_{R}$, $ N^{2}_{z} $, $ N^{2}_\mathrm{eff} $ 
normalized by $ \Omega^{2}_{K} $ and also $ \Delta_{lS} $ normalized by 
$ v^{2}_{\scriptscriptstyle K} $. We can see that $ N^{2}_{R}$ is always negative, 
$ N^{2}_{z} $ is very small and near zero, while $ \kappa^{2} $ is large and positive. 
Consequently, $ N^{2}_\mathrm{eff} $ mostly follows $\kappa^{2} $ pattern and is positive 
in all $ \theta $ angles. As it is seen $ \Delta_{lS} $ is negative based on our
calculation. Since, both Solberg-H\o iland criterions are not satisfy here, 
we conclude that the disc is convectively unstable. This
result is then valid in the absence of magnetic field since numerical MHD 
simulations of accretion flow reveals that flow is convectively 
stable over most of the accretion flow (\citealt{Narayan et al. 2012}; \citealt{Yuan et al. 2012a}).

\subsection{Energy advection} \label{subsec:advection}

To study the energetics of the supercritical accretion flow, we define the vertically averaged
advection parameter, $ f_\mathrm{adv} $, as follows,

\begin{equation}
	f_\mathrm{adv} \equiv \frac{Q_\mathrm{adv}}{Q_\mathrm{vis}} = 1 - \frac{Q_\mathrm{rad}^{-}}{Q_\mathrm{vis}},
\end{equation}
where $ Q_\mathrm{adv} $ is the energy advection rate, 
$ Q_\mathrm{vis} $ is the viscous dissipative heating rate, and $ Q_\mathrm{rad}^{-} $
is the radiative cooling rate. Based on our numerical results, the radiative cooling rate
becomes larger than the viscous dissipation rate at high latitudes. Therefore, the 
$ f_\mathrm{adv} $ becomes negative and plays a heating role rather than cooling one. 
To calculate the advection parameter, we average this quantity over angles  
$ \theta \ge 80^{\circ} $  at $ r = 10 r_\mathrm{s} $ (very 
close to the equator). In this range the advection parameter is positive in our self-similar solution. 
The vertical average of the cooling/heating rates can be written as,

\begin{equation}
	Q_\mathrm{adv} = 2 \int^{90^{\circ}}_{80^{\circ}} q_\mathrm{adv} r \sin \theta \mathrm{d} \theta, 
\end{equation}
\begin{equation}
	Q_\mathrm{vis} = 2 \int^{90^{\circ}}_{80^{\circ}} \Phi_\mathrm{vis} r \sin \theta \mathrm{d} \theta, 
\end{equation}
\begin{equation}
	Q_\mathrm{rad}^{-} = 2 \int^{90^{\circ}}_{80^{\circ}} \bm{\nabla} \cdot \bm{F}\, r \sin \theta \mathrm{d} \theta, 
\end{equation}
where $ q_\mathrm{adv} $ and $ \Phi_\mathrm{vis} $ are the terms presented in 
equation [\ref{ode_energy}]. We found that $ f_\mathrm{adv} \sim 0.62 $ which is
also consistent with the numerical results presented in \citealt{Yang et al. 2014}.

 \section{Summary and Discussion} \label{sec:summary}

We solved two-dimensional RHD equations
of supercritical accretion flows in spherical coordinates and 
in the full $ r-\theta $ space. Our calculations start from 
the rotation axis to the equatorial plane. We adopted the Newtonian potential
for the central black hole. We considered three components of the velocity and 
used $ \alpha $ prescription of the viscosity. We supposed the 
radiation pressure is much more greater than the gas 
pressure, i.e., $ p_\mathrm{gas} / p_\mathrm{rad} \ll 1 $.
Consequently, the gas pressure and also the internal energy density 
of the gas was neglected in our calculations. By adopting the self-similar solutions,
we solved the ODE equations as two point value problem and obtained the variations of
the physical quantities, $ v_r(\theta), v_{\theta}(\theta), v_{\phi}(\theta), \rho(\theta) $,
and $ p(\theta) $ in the vertical direction. We found inflow-outflow solution. 
Similar to our previous work, \citealt{Zeraatgari et al. 2016}, we found that the density profile 
can be described by $ \rho (r) \propto r^{-0.5} $. Correspondingly, the radial profile of 
the mass inflow rate decreases with decreasing radii as $ \dot{M}_\mathrm{in} \propto r $. 
This result is fully consistent with recent analytical and numerical predictions of accretion discs
(e.g., \citealt{Ohsuga et al. 2005}; \citealt{Yuan et al. 2012a, Yuan et al. 2012b}; 
\citealt{Begelman 2012}; \citealt{Yuan et al. 2015}; \citealt{Mosallanezhad et al. 2016}; 
\citealt{Mosallanezhad et al. 2019}). Our results showed
the radiation pressure and the density drop 
from the equatorial plane to the rotation axis.
In the region $ \theta > 65^{\circ} $, there exists inflow
and above that flow moves outward and wind would launch.
Our results also show that there is no sonic point above the disc.
In the supercritical case, we studied here, the radiation could 
push the gas particles outwards and launch the wind with 
poloidal velocity $ \sim 0.25 c $. These results are consistent with 
previous simulations of \citealt{Yang et al. 2014} and \citealt{Ohsuga et al. 2005}.
From our results, the temperature would drop in the wind region
and this clearly shows that the wind produced by the radiation can effectively cool the gas.
By our calculations, the mass flux weighted angular momentum 
of the inflow is lower than that of the wind so the angular momentum 
of the flow can be transported by the wind.
This result is again consistent with previous numerical simulations.
One of our purposes here is to study which force can produce wind in supercritical flow.
Our results show the radial component of the radiation force is the prominent force to drive outflow.
We approximated the convective instability in this study. We found, unlike previous 
analytical works, two Solberg-H\o iland criterions were
not satisfy here, so the disc is convectively unstable in the absence of the magnetic field. 

There are some caveats in this work which we postpone them 
to our future studies. One is that we assume that the gas pressure
is much more lower than the radiation pressure which is not physical. In principle, 
the gas and also the radiation pressures should be comparable with each other.
Another caveat here is that, the total opacity should include both 
absorption and scattering opacities. To avoid complexity, we neglected free-free absorption 
and bound-free absorption in the present study. Moreover, in accretion disc models, 
the magnetic field would be important to transfer angular momentum outward. 
In fact, the inclusion of magnetic field will enhance the outflow as well. 
Therefore, in terms of supercritical disc, it would be interesting to investigate 
the flow structure by combining both the radiation and the magnetic field.

\section*{Aknowledgments}
We thank Feng Yuan for his thoughtful and constructive comments.
Amin Mosallanezhad is supported by the Chinese Academy of Sciences 
President's International Fellowship Initiative, (PIFI), Grant No. 2018PM0046.
This work is supported by National Natural Science Foundation of China (Grant No. 11725312, 11421303) 
and Science Challenge Project of China (Grant No. TZ2016002).

\appendix

\section{steady state and axisymmetric equations in spherical polar coordinates}  \label{appendix_A}

To simplify the equations (\ref{eq:continuity})-(\ref{eq:energy_rad}), we work in spherical coordinates, 
$ (r, \theta, \phi) $. We assume axisymmetric, $ \partial / \partial \phi $ and steady state, $ \partial / \partial t $,
flow and consider all three components of the velocity as $ v_{r} $, $ v_{\theta} $, $ v_{\phi} $. We further 
assume the accretion disc is radiation supported, i.e., the gas pressure is negligible compared to the radiation 
pressure, i.e., $ p_\mathrm{gas} / p_\mathrm{rad} \ll 1 $. Therefore, the gas pressure and the internal energy 
density of the gas are dropped in our equations. Following \citealt{Mihalas and Mihalas 1984}, the components of the viscous 
stress tensor in spherical coordinates are given by,

\begin{equation} \label{sigma_rr}
	\sigma_{rr} = 2 \mu  \left( \frac{\partial v_{r}}{\partial r}  \right) - \frac{2}{3} \mu \left[ \frac{1}{r^{2}} 
	\frac{\partial}{\partial r} \left(r^{2} v_{r} \right) + \frac{1}{r \sin\theta}\frac{\partial}{\partial \theta} 
	\left( v_{\theta} \sin\theta \right) \right] 
\end{equation}

\begin{equation} \label{sigma_tt}
	\sigma_{\theta \theta} = 2 \mu  \left(\frac{v_{r}}{r} +  \frac{1}{r} \frac{\partial v_{\theta}}{\partial \theta}  
	\right) - \frac{2}{3} \mu \left[ \frac{1}{r^{2}} \frac{\partial}{\partial r} \left(r^{2} v_{r} \right) + \frac{1}{r \sin\theta}
	\frac{\partial}{\partial \theta} \left( v_{\theta} \sin\theta \right) \right] 
\end{equation}

\begin{equation} \label{sigma_pp}
	\sigma_{\phi \phi} = 2 \mu  \left(\frac{v_{r}}{r} +  \frac{v_{\theta} \cot \theta}{r}  
	\right) - \frac{2}{3} \mu \left[ \frac{1}{r^{2}} \frac{\partial}{\partial r} \left(r^{2} v_{r} \right) + \frac{1}{r \sin\theta}
	\frac{\partial}{\partial \theta} \left( v_{\theta} \sin\theta \right) \right] 
\end{equation}

\begin{equation} \label{sigma_rt}
	\sigma_{r \theta} = \sigma_{\theta r} = \mu  \left[ \frac{1}{r} \frac{\partial v_{r}}{\partial \theta} 
	+ r \frac{\partial}{\partial r} \left(\frac{v_{\theta}}{r} \right)  \right] 
\end{equation}

\begin{equation} \label{sigma_rp}
	\sigma_{r \phi} = \sigma_{\phi r} = \mu r \frac{\partial}{\partial r} \left(\frac{v_{\phi}}{r} \right) 
\end{equation}

\begin{equation} \label{sigma_tp}
	\sigma_{\theta \phi} = \sigma_{\phi \theta} = \mu  \left[ \frac{\sin \theta}{r}  \frac{\partial}{\partial \theta}  
	\left( \frac{v_{\phi}}{\sin \theta} \right)  \right] 
\end{equation}

By substituting the above equations and considering assumptions described in section 
\ref{sec:equations} the basic equations take the form:

\begin{equation}\label{continuity}
	\frac{1}{r^{2}} \frac{\partial}{\partial r} \left( r^{2} \rho v_{r} \right) + \frac{1}{r \sin  \theta } 
	\frac{\partial}{\partial \theta} \left( \sin \theta  \rho v_{\theta} \right) = 0,
\end{equation}

\begin{multline}\label{momentun1}
	\rho \left[  v_{r} \frac{\partial v_{r}}{\partial r} +\frac{v_{\theta}}{r} \left( \frac{\partial v_{r}}{\partial \theta} 
	- v_{\theta}\right)- \frac{v_{\phi}^{2}}{r} \right] = - \rho \frac{GM}{r^{2}}  + \frac{\rho \kappa_\mathrm{es}}{c} F_{r} \\
	+ \frac{\partial}{\partial r} \Bigg\{ 2\mu \frac{\partial v_{r}}{\partial r} -\frac{2}{3} \mu \left[ \frac{1}{r^{2}} 
	\frac{\partial}{\partial r} \left(r^{2} v_{r} \right) + 
	\frac{1}{r \sin\theta}\frac{\partial}{\partial \theta} 
	\left( v_{\theta} \sin\theta \right) \right] \Bigg\} \\
   	+ \frac{1}{r} \frac{\partial}{\partial \theta} \Bigg\{ \mu \left[ r \frac{\partial}{\partial r} \left( \frac{v_{\theta}}{r} \right) 
	+ \frac{1}{r} \frac{\partial v_{r}}{\partial \theta} \right] \Bigg\} + \frac{\mu}{r} \left[ 4 r \frac{\partial}{\partial r} 
	\left( \frac{v_{r}}{r} \right) - \frac{2}{r \sin \theta} \frac{\partial}{\partial \theta} \left( v_{\theta} \sin \theta \right) 
	+ r \cot \theta \frac{\partial}{\partial r} \left( \frac{v_{\theta}}{r}\right) + \frac{\cot \theta}{r} \frac{\partial v_{r}}{\partial \theta} \right],
\end{multline}

\begin{multline}\label{momentum2}
	\rho \left[  v_{r} \frac{\partial v_{\theta}}{\partial r} +\frac{v_{\theta}}{r} \left( \frac{\partial v_{\theta}}{\partial \theta} 
	+ v_{r}\right)- \frac{v_{\phi}^{2}}{r} \cot \theta \right] =  \frac{\rho \kappa_\mathrm{es}}{c} F_{\theta} 
	+ \frac{\partial}{\partial r} \Bigg\{  \mu \left[  r \frac{\partial}{\partial r} \left( \frac{v_{\theta}}{r} \right) 
	+ \frac{1}{r} \frac{\partial v_{r}}{\partial \theta}  \right]  \Bigg\} \\
	+ \frac{1}{r} \frac{\partial}{\partial \theta} \Bigg\{ \frac{2\mu}{r} \left( \frac{\partial v_{\theta}}{\partial \theta} 
	+ v_{r} \right) - \frac{2}{3} \mu \left[ \frac{1}{r^{2}} \frac{\partial}{\partial r} \left(r^{2} v_{r} \right) 
	+ \frac{1}{r \sin\theta}\frac{\partial}{\partial \theta} \left( v_{\theta} \sin\theta \right) \right] \Bigg\} \\
	+ \frac{\mu}{r} \Bigg\{  \frac{2 \cot \theta}{r} \left[ \sin \theta \frac{\partial}{\partial \theta} 
	\left( \frac{v_{\theta}}{\sin \theta} \right) \right] + 3 r \frac{\partial}{\partial r} \left( \frac{v_{\theta}}{r} \right) 
	+ \frac{3}{r} \frac{\partial v_{r}}{\partial \theta}  \Bigg\},
\end{multline}

\begin{multline}\label{momentum3}
	\rho \left[ v_{r} \frac{\partial v_{\phi}}{\partial r} + \frac{v_{\theta}}{r} \frac{\partial v_{\phi}}{\partial \theta} 
	+ \frac{v_{\phi}}{r} \left( v_{r} +  v_{\theta} \cot\theta \right)  \right] = 
 	\frac{\partial}{\partial r} \left[  \mu r \frac{\partial}{\partial r} \left(  \frac{v_{\phi}}{r} \right)  \right]  
	+ \frac{1}{r}  \frac{\partial}{\partial \theta} \left[  \frac{\mu \sin \theta}{r} \frac{\partial}{\partial \theta}  
	\left( \frac{v_{\phi}}{\sin \theta} \right)  \right] \\
	+ \frac{\mu}{r} \left[ 3 r \frac{\partial}{\partial r} \left( \frac{v_{\phi}}{r} \right) + \frac{2 \cos\theta}{r} 
	\frac{\partial}{\partial \theta} \left( \frac{v_{\phi}}{\sin \theta} \right) \right],
\end{multline}

\begin{multline}\label{energy}
	\frac{1}{r^{2}} \frac{\partial}{\partial r} \left( r^{2} E v_{r} \right) + \frac{1}{r \sin  \theta } 
	\frac{\partial}{\partial \theta} \left( \sin \theta  E v_{\theta} \right)  + \frac{E}{3}  
	\left[ \frac{\partial v_{r}}{\partial r} + \frac{1}{r} \left( \frac{\partial v_{\theta}}{\partial \theta}  
	+ v_{r}\right) + \frac{1}{r} \left( v_{r} + v_{\theta} \cot \theta \right)  \right]  = 
	 - \frac{1}{r^{2}} \frac{\partial }{\partial r} \left( r^{2}\, F_{r} \right) \\ 
	- \frac{1}{r \sin\theta} \frac{\partial}{\partial \theta} \left( \sin\theta\, F_{\theta} \right) 
	+ 2 \mu \Bigg\{  \left( \frac{\partial v_{r}}{\partial r} \right)^{2}  +  \left( \frac{1}{r} 
	\frac{\partial v_{\theta}}{\partial \theta}  + \frac{v_{r}}{r}  \right)^{2}  
	+  \left(  \frac{v_{r}}{r} + \frac{v_{\theta} \cot \theta}{r} \right)^{2} + \frac{1}{2} 
	\left[ r \frac{\partial}{\partial r} \left( \frac{v_{\theta}}{r} \right)  + \frac{1}{r} 
	\frac{\partial v_{r}}{\partial \theta}  \right]^{2}  \\
	+ \frac{1}{2} \left[ r \frac{\partial}{\partial r} \left( \frac{v_{\phi}}{r} \right) \right]^{2} 
	+ \frac{1}{2} \left[  \frac{\sin \theta}{r} \frac{\partial}{\partial \theta} \left(  \frac{v_{\phi}}{\sin \theta}  \right)   
	\right]^{2} \Bigg\} - \frac{2}{3} \mu \left[ \frac{1}{r^{2}} \frac{\partial}{\partial r} \left(r^{2} v_{r} \right) 
	+ \frac{1}{r \sin\theta}\frac{\partial}{\partial \theta} \left( v_{\theta} \sin\theta \right) \right]^{2},
\end{multline}
where

\begin{equation} \label{F_r}
	F_{r} = - \frac{c}{3 \rho \kappa_\mathrm{es}} \frac{\partial E}{\partial r},
\end{equation}

\begin{equation} \label{F_t}
	F_{\theta} = - \frac{c}{3 \rho \kappa_\mathrm{es}} \frac{1}{r} \frac{\partial E}{\partial \theta}.
\end{equation}

\section{Ordinary Differential Equations}  \label{appendix_B}

By applying the equations (\ref{self_similar_vr})-(\ref{self_similar_p}) into partial differential 
equations in Appendix \ref{appendix_A} (equations [\ref{continuity}]-[\ref{energy}]), we obtain following 
five coupled ordinary differential equations in $ \theta $ direction as,

\begin{equation} \label{ode_continuity}
	\rho \left[  v_{r} + v_{\theta} \cot \theta + \frac{\mathrm{d} v_{\theta} }{\mathrm{d} \theta} \right] 
	+ v_{\theta} \frac{\mathrm{d} \rho}{\mathrm{d} \theta} = 0
\end{equation}

\begin{multline} \label{ode_mom1}
	\rho \left[ -\frac{1}{2} v_{r}^2 + v_{\theta}\frac{\mathrm{d} v_{r}}{\mathrm{d} \theta} - v_{\theta}^2 
	- \Omega^2 \sin^2 \theta \right] = - \rho  + \frac{3}{2} p + \alpha p \frac{\mathrm{d}^2 v_{r}}{\mathrm{d} \theta^2} \\
	+ \alpha \Bigg\{ \left[ - 3 v_{r}  + \frac{\mathrm{d} v_{r}}{\mathrm{d} \theta}  \cot \theta - \frac{5}{2} 
	\left( \frac{\mathrm{d} v_{\theta}}{\mathrm{d} \theta} +  v_{\theta} \cot \theta \right) \right] p  
	+ \left[ \frac{\mathrm{d} v_{r}}{\mathrm{d} \theta} - \frac{3}{2} v_{\theta} \right] \frac{\mathrm{d} p}{\mathrm{d} \theta} \Bigg\}
\end{multline}

\begin{multline} \label{ode_mom2}
	\rho \left[ \frac{1}{2} v_{r} v_{\theta} + v_{\theta} \frac{dv_{\theta}}{d \theta} 
	- \cos \theta \sin \theta \Omega^{2} \right] = - \frac{d p}{d \theta} 
	+ \frac{4}{3} \alpha  p \frac{\mathrm{d}^{2} v_{\theta}}{\mathrm{d} \theta^{2}} \\
	+ \alpha \Bigg\{ \left[ \frac{5}{2} \frac{\mathrm{d}v_{r}}{\mathrm{d} \theta} 
	- v_{\theta} \left( \frac{9}{4} + \frac{4}{3} \cot^{2} \theta \right) 
	+ \frac{4}{3} \frac{\mathrm{d}v_{\theta}}{\mathrm{d} \theta} \cot \theta \right] p 
	+ \left[ v_{r} + \frac{4}{3} \frac{\mathrm{d} v_{\theta}}{\mathrm{d} \theta} 
	- \frac{2}{3} v_{\theta} \cot \theta \right] \frac{\mathrm{d} p}{\mathrm{d} \theta} \Bigg\}
\end{multline}

\begin{equation} \label{ode_mom3}
	\rho \left[ \frac{1}{2} v_{r} \Omega \sin \theta 
	+ v_{\theta} \frac{\mathrm{d} \Omega}{\mathrm{d} \theta} \sin \theta 
	+  2 v_{\theta} \Omega \cos \theta \right] = 
	\alpha \Bigg\{ - \frac{9}{4} p \Omega \sin \theta 
	+  p \frac{\mathrm{d}^2 \Omega}{\mathrm{d} \theta^2} \sin \theta 
	+ \frac{\mathrm{d} p}{\mathrm{d} \theta}  \frac{\mathrm{d} \Omega}{\mathrm{d} \theta} \sin \theta 
	+ 3 p \frac{\mathrm{d} \Omega}{\mathrm{d} \theta} \cos \theta \Bigg\}
\end{equation}

\begin{multline} \label{ode_energy}
	\frac{3}{2}  v_{r} p + 3 v_{\theta} \frac{\mathrm{d}p}{\mathrm{d} \theta} 
	+ 4 p \frac{\mathrm{d} v_{\theta}}{\mathrm{d} \theta} + 4 v_{\theta} p \cot \theta 
	= \tau_{0}^{-1} \left( \frac{c}{v_{0}} \right)   
	\times \left[ - \frac{1}{\rho^{2}} \frac{\mathrm{d} p}{\mathrm{d} \theta} \frac{\mathrm{d} \rho}{\mathrm{d} \theta} 
	+ \frac{1}{\rho} \frac{\mathrm{d}^2 p }{\mathrm{d} \theta^2} 
	+ \frac{1}{\rho} \frac{\mathrm{d} p}{\mathrm{d} \theta} \cot \theta  \right] \\
	+ \alpha p \Bigg\{ \frac{1}{2} v_{r}^{2} + 2 \left( v_{r} + \frac{\mathrm{d} v_{\theta}}{\mathrm{d} \theta} \right)^{2} 
	+ 2 \left( v_{r} + v_{\theta} \cot \theta \right)^{2} + \frac{1}{4} \left( 3 v_{\theta} 
	- 2 \frac{\mathrm{d} v_{r}}{\mathrm{d} \theta} \right)^{2} 
	+ \sin^{2} \theta \left[ \frac{9}{4} \Omega^{2} 
	+ \left( \frac{d \Omega}{d \theta} \right)^{2} \right] \\
	- \frac{2}{3} \left[  \frac{3}{2} v_{r} + \frac{\mathrm{d} v_{\theta}}{\mathrm{d} \theta} 
	+ v_{\theta} \cot \theta \right]^{2} \Bigg\}
\end{multline}
where $ \tau_{0} (= \rho_{0} r_{0} \kappa_\mathrm{es}) $ is the midplane optical depth at $ r_{0} $.
The above ODEs represent the variation of five scaler quantities, $ v_{r} (\theta) $, $ v_{\theta}(\theta) $, 
$ v_{\phi}(\theta) $, $ \rho(\theta) $, and $ p(\theta) $ in $ \theta $ (for simplicity, we remove the $ \theta $ 
dependency of our physical variables in the above equations).


\begin{thebibliography}{}
 
\bibitem[\protect\citeauthoryear{Abramowicz et al.}{1988}]{Abramowicz et al. 1988} Abramowicz, M.~A., Czerny, B., Lasota, J.~P., Szuszkiewicz, E.\ 1988, \apj, 332, 646

\bibitem[\protect\citeauthoryear{Abramowicz \& Fragile}{2013}]{Abramowicz and Fragile 2013} Abramowicz, M.~A., Fragile, P.~C.\ 2013, Living Rev. Relativ. , 16,1 
\bibitem[\protect\citeauthoryear{Balbus \& Hawley}{1998}]{Balbus and Hawley 1998} Balbus, S. A., Hawley, J. F.\ 1998, \rmp, 70, 1
\bibitem[Begelman \& Meier(1982)]{Begelman and Meier 1982} Begelman, M.~C., Meier, D.~L.\ 1982, \apj, 253, 873
\bibitem[\protect\citeauthoryear{Begelman \& Meier}{1982}]{Begelman and Meier 1982} Begelman, M.~C., Meier, D.~L.\ 1982, \apj, 253, 873
\bibitem[\protect\citeauthoryear{Begelman}{2012}]{Begelman 2012} Begelman, M.~C.\ 2012, \apj, 749, 3
\bibitem[\protect\citeauthoryear{Blaes}{2014}]{Blaes 2014} Blaes, O. 2014,  \ssr, 183, 21
\bibitem[\protect\citeauthoryear{Blandford \& Begelman}{1999}]{Blandford and Begelman 1999} Blandford, R. D., \& Begelman, M. 1999, MNRAS, 303, 1 
\bibitem[\protect\citeauthoryear{Blandford \& Begelman}{2004}]{Blandford and Begelman 2004} Blandford, R. D., \& Begelman, M. 2004, MNRAS, 349, 68
\bibitem[\protect\citeauthoryear{Boller}{2000}]{Boller 2000} Boller, T.\ 2000, \nar, 44, 387
\bibitem[\protect\citeauthoryear{Done et al.}{2007}]{Done et al. 2007} Done, C., Gierlinski, M., \& Kubota, A.\ 2007, \aapr , 15, 1
\bibitem[\protect\citeauthoryear{Eggum et al.}{1988}]{Eggum et al. 1988} Eggum, G. E., Coroniti, F. V., \& Katz, J. I.\ 1988, \apj, 330, 142
\bibitem[\protect\citeauthoryear{Eggum et al.}{1987}]{Eggum et al. 1987} Eggum, G.~E., Coroniti, F.~V., \& Katz, J.~I.\ 1987, \apj, 323, 634
\bibitem[\protect\citeauthoryear{Fragile et al.}{2014}]{Fragile et al. 2014} Fragile, P. C., Olejar, A., \& Anninos, P.\ 2014, \apj, 796, 22
\bibitem[\protect\citeauthoryear{Frank et al.}{2002}]{Frank et al. 2002} Frank, J., King, A., \& Raine, D.~J.\ 2002. Accretion Power in Astrophysics. Cambridge, UK: Cambridge Univ. Press
\bibitem[\protect\citeauthoryear{F\"{u}rst et al.}{2016}]{Furst et al. 2016} F\"{u}rst, F., Walton, D.~J., Harrison, F.~A., et al.\ 2016, \apjl, 831, L14
\bibitem[\protect\citeauthoryear{Fukue}{2004}]{Fukue 2004} Fukue, J.\ 2004, \pasj, 56, 569
\bibitem[\protect\citeauthoryear{Gu \& Lu}{2007}]{Gu and Lu 2007} Gu, W.~M., \& Lu, J.~F.\ 2007, \apj, 660, 541
\bibitem[\protect\citeauthoryear{Gu}{2012}]{Gu 2012} Gu, W.~M.\ 2012, \apj, 753, 118
\bibitem[\protect\citeauthoryear{Israel et al.}{2017a}]{Israel et al. 2017a} Israel, G. L., Belfiore, A., Stella, L., et al.\ 2017a, Sci., 355, 817
\bibitem[Israel et al.(2017b)]{} Israel, G. L., Papitto, A., Esposito, P., et al.\ 2017b, \mnras, 466, L48
\bibitem[\protect\citeauthoryear{Israel et al.}{2017b}]{Israel et al. 2017b} Israel, G. L., Papitto, A., Esposito, P., et al.\ 2017b, \mnras, 466, L48
\bibitem[\protect\citeauthoryear{Jiang et al.}{2014}]{Jiang et al. 2014} Jiang, Y.~F., Stone, J. M., \& Davis, S. W.\ 2014, \apj, 796, 106
\bibitem[\protect\citeauthoryear{Kaaret et al.}{2017}]{Kaaret et al. 2017} Kaaret, P., Feng, H., \& Roberts, T. P.\ 2017, \araa, 55, 303
\bibitem[\protect\citeauthoryear{Kato et al.}{2008}]{Kato et al. 2008} Kato, S., Fukue, J., Mineshige, S.\ 2008. Black-Hole Accretion Disks: Towards a New Paradigm. Kyoto: Kyoto Univ. Press
\bibitem[\protect\citeauthoryear{Kawashima et al.}{2009}]{Kawashima et al. 2009} Kawashima, T., Ohsuga, K., Mineshige, S., et al. 2009, \pasj, 61, 769
\bibitem[\protect\citeauthoryear{Kawashima et al.}{2012}]{Kawashima et al. 2012} Kawashima T., Ohsuga K., Mineshige S., et al.\ 2012, \apj, 752, 18
\bibitem[\protect\citeauthoryear{Kitaki et al.}{2017}]{Kitaki et al. 2017} Kitaki, T., Mineshige, S., Ohsuga, K., Kawashima, T.\ 2017, \pasj, 69, 92 
\bibitem[\protect\citeauthoryear{Kitaki et al.}{2018}]{Kitaki et al. 2018} Kitaki, T., Mineshige, S., Ohsuga, K., Kawashima, T.\ 2018, \pasj, 70, 108 
\bibitem[\protect\citeauthoryear{Koratkar \& Blaes}{1999}]{Koratkar and Blaes 1999} Koratkar, A., Blaes, O.\ 1999. \pasj, 755, 1
\bibitem[\protect\citeauthoryear{Kosec et al.}{2018}]{Kosec et al. 2018} Kosec, P., Pinto, C., Walton, D.~J., Fabian, A.~C., Bachetti, M., Brightman, M., F\"{u}rst, F., \& Grefenstette, B.~W.\ 2018, \mnras, 479, 3978
\bibitem[\protect\citeauthoryear{Levermore \& Pomraning}{1981}]{Levermore and Pomraning 1981} Levermore, C.~D., Pomraning, G.~C.\ 1981, \apj, 248, 321
\bibitem[\protect\citeauthoryear{Lynden-Bell \& Pringle}{1974}]{Lynden-Bell and Pringle 1974} Lynden-Bell, D., Pringle, J.~E.\ 1974, \mnras, 168:603, 37
\bibitem[\protect\citeauthoryear{Makishima et al.}{2000}]{Makishima et al. 2000} Makishima, K., Kubota, A., Mizuno, T., Ohnishi, T., Tashiro, M., Aruga, Y., Asai, K., Dotani, T., et al.\ 2000, \apj, 535, 632
\bibitem[\protect\citeauthoryear{McKinney et al.}{2014}]{McKinney et al. 2014} McKinney, J. C., Tchekhovskoy, A., Sadowski, A., \& Narayan, R.\ 2014, \mnras, 441, 3177
\bibitem[\protect\citeauthoryear{McClintock et al.}{2014}]{McClintock et al. 2014} McClintock J.~E., Narayan, R., \& Steiner, J.~F.\ 2014, \ssr, 183, 295
\bibitem[\protect\citeauthoryear{Mihalas \& Mihalas}{1984}]{Mihalas and Mihalas 1984} Mihalas D., Mihalas B.~W.\ 1984, Foundations of Radiation Hydrodynamics. Oxford Univ. Press, Oxford
\bibitem[\protect\citeauthoryear{Miller et al.}{2004}]{Miller et al. 2004} Miller, J.~M., Fabian, A.~C. \& Miller, M.~C. \ 2004, \apj, 614, L117
\bibitem[\protect\citeauthoryear{Mineshige et al.}{2000}]{Mineshige et al. 2000} Mineshige, S., Kawaguchi, T., Takeuchi, M., \& Hayashida, K.\ 2000, \pasj, 52, 499
\bibitem[\protect\citeauthoryear{Mosallanezhad et al.}{2014}]{Mosallanezhad et al. 2014}Mosallanezhad, A., Abbassi, S., Beiranvand, N.\ 2014, \mnras, 437, 3112
\bibitem[\protect\citeauthoryear{Mosallanezhad et al.}{2016}]{Mosallanezhad et al. 2016} Mosallanezhad, A., Bu, D.~F., Yuan, F.\ 2016, \mnras, 456, 2877
\bibitem[\protect\citeauthoryear{Mosallanezhad et al.}{2019}]{Mosallanezhad et al. 2019} Mosallanezhad, A., Yuan, F., Ostriker, J., P., et al.\ 	arXiv:1910.00288 
\bibitem[\protect\citeauthoryear{Narayan et al.}{2000}]{Narayan et al. 2000} Narayan, R., Igumenshchev, I. V., \& Abramowicz, M. A. 2000, ApJ, 539, 798
\bibitem[\protect\citeauthoryear{Narayan \& Yi}{1994}]{Narayan and Yi 1994}Narayan, R., Yi, I.\ 1994, \apjl, 428, L13
\bibitem[\protect\citeauthoryear{Narayan \& Yi}{1995}]{Narayan and Yi 1995} Narayan, R., Yi, I.\ 1995, \apj, 444, 231
\bibitem[\protect\citeauthoryear{Narayan et al.}{2012}]{Narayan et al. 2012} Narayan, R., Sadowski, A., Penna, R. F., Kulkarni, A. K.\ 2012, \mnras, 426, 3241
\bibitem[\protect\citeauthoryear{Novikov \& Thorne}{1973}]{Novikov and Thorne 1973} Novikov, I. D., \& Thorne, K. S.\ 1973, in Black Holes, ed. C. DeWitt \& B. DeWitt (New York: Gordon and Breach), 343
\bibitem[\protect\citeauthoryear{Ohsuga \& Mineshige}{2007}]{Ohsuga and Mineshige 2007} Ohsuga, K., \& Mineshige, S.\ 2007, \apj, 670, 1283
\bibitem[\protect\citeauthoryear{Ohsuga \& Mineshige}{2011}]{Ohsuga and Mineshige 2011} Ohsuga K., Mineshige S.\ 2011, \apj, 736, 2 
\bibitem[\protect\citeauthoryear{Ohsuga et al.}{2005}]{Ohsuga et al. 2005} Ohsuga, K., Mori, M., Nakamoto, T., \& Mineshige, S.\ 2005, \apj, 628, 368
\bibitem[\protect\citeauthoryear{Ohsuga et al.}{2009}]{Ohsuga et al. 2009} Ohsuga, K., Mineshige, S., Mori, M., \& Kato, Y.\ 2009, \pasj, 61, L7
\bibitem[\protect\citeauthoryear{Okuda et al.}{2005}]{Okuda et al. 2005}Okuda, T., Teresi, V., Toscano, E., \& Molteni, D.\ 2005, \mnras, 357, 295
\bibitem[\protect\citeauthoryear{Okuda}{2002}]{Okuda 2002} Okuda, T.\ 2002, \pasj, 54, 253
\bibitem[\protect\citeauthoryear{Pringle}{1981}]{Pringle 1981} Pringle, J.~E.\ 1981, \araa, 19:137, 62
\bibitem[\protect\citeauthoryear{Remillard \& McClintock}{2006}]{Remillard and McClintock 2006} Remillard, R.~A., McClintock, J.~E.\ 2006, \araa, 44, 49
\bibitem[\protect\citeauthoryear{Sadowski et al.}{2009}]{Sadowski et al. 2009} Sadowski, A., Abramowicz, M., Bursa, M., et al.\ 2009, \aap, 502, 7 
\bibitem[\protect\citeauthoryear{Sadowski et al.}{2011}]{Sadowski et al. 2011} Sadowski, A., Abramowicz, M., Bursa, M., et al.\ 2011, \aap, 527, A17
\bibitem[\protect\citeauthoryear{Sadowski et al.}{2014}]{Sadowski et al. 2014} Sadowski, A., Narayan, R., McKinney, J.~C., \& Tchekhovskoy, A.\ 2014, \mnras, 439, 503
\bibitem[\protect\citeauthoryear{Sadowski et al.}{2015}]{Sadowski et al. 2015} Sadowski, A., Narayan, R., Tchekhovskoy, A., et al.\ 2015, \mnras, 447, 49
\bibitem[\protect\citeauthoryear{Shakura \& Sunyaev}{1973}]{Shakura and Sunyaev 1973} Shakura, N. I., Sunyaev, R. A.\ 1973, \aap, 24, 337
\bibitem[\protect\citeauthoryear{Stone et al.}{1999}]{Stone et al. 1999} Stone, J. M., Pringle, J. E., \& Begelman, M. C.\ 1999, \mnras, 310, 1002
\bibitem[\protect\citeauthoryear{Takahashi et al.}{2016}]{Takahashi et al. 2016} Takahashi, H.~R., Ohsuga, K., Kawashima, T., \& Sekiguchi, Y.\ 2016, \apj, 826, 23
\bibitem[\protect\citeauthoryear{Vierdayanti et al.}{2010}]{Vierdayanti et al. 2010} Vierdayanti, K., Mineshige, S., \& Ueda, Y.\ 2010, \pasj, 62, 239
\bibitem[\protect\citeauthoryear{Wang \& Zhou}{1999}]{Wang and Zhou 1999}Wang, J.~M., \& Zhou, Y.~Y.\ 1999, \apj, 516, 420
\bibitem[\protect\citeauthoryear{Watarai \& Fukue}{1999}]{Watarai and Fukue 1999} Watarai, K., \& Fukue, J.\ 1999, \pasj, 51, 725
\bibitem[\protect\citeauthoryear{Watarai et al.}{2000}]{Watarai et al. 2000} Watarai, K., Fukue, J., Takeuchi, M., \& Mineshige, S.\ 2000, \pasj, 52, 133
\bibitem[\protect\citeauthoryear{Watarai et al.}{2001}]{Watarai et al. 2001} Watarai, K., Mizuno, T., \& Mineshige, S.\ 2001, \apjl, 549, L77
\bibitem[\protect\citeauthoryear{Watarai}{2006}]{Watarai 2006} Watarai, K.\ 2006, \apj, 648, 523 
\bibitem[\protect\citeauthoryear{Yang et al.}{2014}]{Yang et al. 2014} Yang, X.~H., Yuan, F., Ohsuga, K.,\& Bu, D.\ 2014, 780, 79
\bibitem[\protect\citeauthoryear{Yuan \& Narayan}{2014}]{Yuan and Narayan 2014} Yuan, F., Narayan, R.\ 2014, \araa, 52, 529 
\bibitem[\protect\citeauthoryear{Yuan et al.}{2012a}]{Yuan et al. 2012a} Yuan, F., Bu, D., Wu, M.\ 2012a. \apj, 761, 130
\bibitem[\protect\citeauthoryear{Yuan et al.}{2015}]{Yuan et al. 2015} Yuan, F., Gan, Z. M., Narayan, R., Sadowski, A., Bu, D., Bai, X. N.\ 2015, \apj, 804, 101
\bibitem[\protect\citeauthoryear{Yuan et al.}{2012b}]{Yuan et al. 2012b} Yuan, F., Wu, M., Bu, D.\ 2012b. \apj, 761,129
\bibitem[\protect\citeauthoryear{Zeraatgari \& Abbassi}{2015}]{Zeraatgari and Abbassi 2015} Zeraatgari, F.~Z., Abbassi, S.\ 2015, \apj, 809, 54
\bibitem[\protect\citeauthoryear{Zeraatgari et al.}{2016}]{Zeraatgari et al. 2016} Zeraatgari, F.~Z., Abbassi, S., Mosallanezhad, A.\ 2016, \apj, 823, 92
\bibitem[\protect\citeauthoryear{Zeraatgari et al.}{2018}]{Zeraatgari et al. 2018} Zeraatgari, F.~Z., Mosallanezhad, A., Abbassi, S., Yuan, Y.~F.\ 2018, \apj, 852, 124


%
\end{thebibliography}
\end{document}